\documentclass[journal]{IEEEtran}
\usepackage{cite}
\ifCLASSINFOpdf
\else
\fi
\usepackage{amsmath}
\ifCLASSOPTIONcompsoc
  \usepackage[caption=false,font=normalsize,labelfont=sf,textfont=sf]{subfig}
\else
  \usepackage[caption=false,font=footnotesize]{subfig}
\fi

\usepackage{amssymb}
\usepackage{amsfonts}
\usepackage{booktabs}
\usepackage{multirow}
\usepackage{hyperref}
\usepackage{siunitx}
\usepackage{mathtools,mathrsfs}
\usepackage{comment}
\usepackage[linesnumbered,ruled,vlined]{algorithm2e}
\usepackage{color}
\usepackage{dsfont}

\def\equationautorefname~#1\null{(#1\null)}

\let\oldnl\nl% Store \nl in \oldnl
\newcommand{\nonl}{\renewcommand{\nl}{\let\nl\oldnl}}% Remove line number for one line
\newlength\mylen
\newcommand\myinput[1]{%
  \settowidth\mylen{\KwOut{}}%
  \setlength\hangindent{\mylen}%
  \nonl\hspace*{\mylen}#1}

\SetCommentSty{mycommfont}

\usepackage{etoolbox}
\makeatletter
\patchcmd{\@algocf@start}% <cmd>
  {-1.5em}% <search>
  {0pt}% <replace>
  {}{}% <success><failure>
\makeatother

\usepackage{xspace}
\makeatletter
\DeclareRobustCommand\onedot{\futurelet\@let@token\@onedot}
\def\@onedot{\ifx\@let@token.\else.\null\fi\xspace}
\def\eg{e.g\onedot} 
\def\ie{i.e\onedot}

\def\etal{et al\onedot}
\makeatother

\newcommand{\vect}[1]{\mbox{\boldmath $#1$}}
\newcommand{\abs}[1]{\left\lvert#1\right\rvert}
\newcommand{\norm}[1]{\left\lVert#1\right\rVert}

\newcommand{\trans}[1]{#1^\mathsf{T}}

\DeclareMathOperator*{\argmax}{arg\,max}
\def\appendixautorefname~#1\null{~#1 \null}
\newcommand{\relmid}{\mathrel{}\middle|\mathrel{}}

\makeatletter
\newcommand{\figcaption}[1]{\def\@captype{figure}\caption{#1}}
\newcommand{\tblcaption}[1]{\def\@captype{table}\caption{#1}}
\makeatother

\begin{document}
\title{Encoder-Decoder Based Attractors\\for End-to-End Neural Diarization}

\author{Shota Horiguchi,~\IEEEmembership{Member,~IEEE,}
        Yusuke Fujita,~\IEEEmembership{Member,~IEEE,}
        Shinji Watanabe,~\IEEEmembership{Senior Member,~IEEE,}\\
        Yawen Xue,~\IEEEmembership{}
        Paola Garc\'{i}a,~\IEEEmembership{Member,~IEEE,}% <-this % stops a space
\thanks{S. Horiguchi and Y. Xue are with Hitachi, Ltd.}%
\thanks{Y. Fujita is with LINE corporation. This work had been done during he was with Hitachi, Ltd.}% <-this % stops a space
\thanks{S. Watanabe is with Carnegie Mellon University.}%
\thanks{P. Garc\'{i}a is with Johns Hopkins University.}%
\thanks{\copyright 2022 IEEE. Personal use of this material is permitted. Permission from IEEE must be obtained for all other uses, in any current or future media, including reprinting/republishing this material for advertising or promotional purposes, creating new collective works, for resale or redistribution to servers or lists, or reuse of any copyrighted component of this work in other works.}%
%\thanks{Manuscript received June 20, 2021; revised January 27, 2022; accepted March 8, 2022.}%
}


\markboth{Journal of \LaTeX\ Class Files,~Vol.~14, No.~8, August~2015}%
{Shell \MakeLowercase{\textit{et al.}}: Bare Demo of IEEEtran.cls for IEEE Journals}



\maketitle

\begin{abstract}
This paper investigates an end-to-end neural diarization (EEND) method for an unknown number of speakers.
In contrast to the conventional cascaded approach to speaker diarization, EEND methods are better in terms of speaker overlap handling.
However, EEND still has a disadvantage in that it cannot deal with a flexible number of speakers.
To remedy this problem, we introduce encoder-decoder-based attractor calculation module (EDA) to EEND.
Once frame-wise embeddings are obtained, EDA sequentially generates speaker-wise attractors on the basis of a sequence-to-sequence method using an LSTM encoder-decoder.
The attractor generation continues until a stopping condition is satisfied; thus, the number of attractors can be flexible.
Diarization results are then estimated as dot products of the attractors and embeddings.
The embeddings from speaker overlaps result in larger dot product values with multiple attractors; thus, this method can deal with speaker overlaps.
Because the maximum number of output speakers is still limited by the training set, we also propose an iterative inference method to remove this restriction. 
Further, we propose a method that aligns the estimated diarization results with the results of an external speech activity detector, which enables fair comparison against cascaded approaches.
Extensive evaluations on simulated and real datasets show that EEND-EDA outperforms the conventional cascaded approach.
\end{abstract}


% \fi
\IEEEpeerreviewmaketitle

\section{Introduction}
\IEEEPARstart{S}{peaker} diarization is a task of estimating multiple speakers' speech activities from input audio (sometimes referred to as the ``who spoke when'' problem) \cite{park2022review}.
It can be placed as a downstream task of automatic speech recognition (ASR), in which speaker information is tagged to each transcribed utterance \cite{carletta2007unleashing,watanabe2020chime,chen2020continuous}.
It can also be used as a prior step to speech separation and the following ASR.
For example, in guided source separation \cite{boeddeker2018front}, speech activities are used as constraints to update time-frequency masks of a complex angular central Gaussian mixture model.
The speech-activity-driven speech-extraction neural network \cite{delcroix2021speaker} takes acoustic features and a target speaker's speech activity to perform fully neural speech separation.

Classical cascaded methods treat speaker diarization as a partition problem.
Given a set of time frames, they first detect speaker-active frames and then divide them into clusters by using speaker embeddings extracted with a sliding window.
The number of clusters, which represents the number of speakers, is determined in the clustering step during inference.
Eigen value analysis on the graph Laplacian of a similarity matrix calculated from frame-wise embeddings is one way to estimate the number of speakers explicitly \cite{wang2018speaker,park2020auto}.
If agglomerative hierarchical clustering is employed as a clustering algorithm, a threshold value is usually preset, and the number of clusters, \ie, the number of speakers, is dynamically determined by the threshold value \cite{sell2014speaker}.
Either way, the number of clusters can be set flexibly during inference.
However, there is one fundamental problem that it basically cannot handle speaker overlaps because each speech frame is usually assigned to one speaker.

Some neural-network-based end-to-end methods, in comparison, naturally handle speaker overlap with a single network.
For example, the Recurrent Selective Attention Network (RSAN) \cite{kinoshita2018listening,von2019all} decodes speech activity for each speaker one by one until a stopping condition is satisfied.
However, it requires clean speech to be trained as a mask-based speech separation model.
End-to-end neural diarization (EEND) \cite{fujita2019end1,fujita2019end2,fujita2020endtoend}, which estimates multiple speakers' speech activities at once from input audio, does not require such clean speech for training.
The limitation is that the original EEND fixes the output number of speakers; thus, knowing the number of speakers in advance is a requirement.

In our previous study \cite{horiguchi2020endtoend}, we introduced an encoder-decoder-based attractor calculation module (EDA) as part of the self-attentive EEND model \cite{fujita2019end2} to handle unknown numbers of speakers (EEND-EDA).
It calculates attractors from frame-wise embeddings using a sequence-to-sequence method with an LSTM encoder-decoder; thus, the number of attractors can be flexible.
In general, sequence-to-sequence methods require a stopping criterion in their decoding process.
To decide when to stop the attractor calculation, EDA also estimates whether each calculated attractor really corresponds to a speaker.
The diarization results are calculated as dot products between the attractors and frame-wise embeddings.
Despite being designed for the diarization of flexible numbers of speakers, it also has performed better than the original EEND under fixed-number-of-speakers conditions.
Compared with other EEND extensions for unknown numbers of speakers \cite{fujita2020neural,takashima2021endtoend}, it performed the best on various datasets including the CALLHOME and DIHARD III datasets \cite{horiguchi2021hitachi}.
Several studies have also proposed extensions to EEND-EDA to allow online processing \cite{han2021bwedaeend,xue2021online2}.

In this paper, we revisit EEND-EDA with more comprehensive discussions and formulations and propose several extensions from the original EEND-EDA presented in \cite{horiguchi2020endtoend}.
The modifications from the original EEND-EDA study are summarized as follows:
\begin{itemize}
    \item We discuss the relationship between the original EEND and EEND-EDA, which explains EEND-EDA's better performance in a fixed-number-of-speakers evaluation.
    \item We also propose refining the training strategy of EEND-EDA, which resulted in a \SI{2.41}{\percent} DER improvement on the CALLHOME dataset from the original paper \cite{horiguchi2020endtoend}.
    \item In the history of diarization studies, it has been difficult to compare the results of cascaded approaches and EEND-based approaches because the former ones are often evaluated with an oracle speech activity detection (SAD), while EENDs operate SAD and diarization simultaneously. To conduct fair comparisons between cascaded and EEND-based approaches, this paper introduces SAD post-processing to align diarization results from EEND-EDA with external SAD results.
    \item We also propose an iterative inference for handling the problem of the number of outputs of EEND-EDA being empirically limited by its training dataset.
    \item We conduct thorough evaluations and analyses on simulated and real datasets including CALLHOME, CSJ, AMI, DIHARD II, and DIHARD III.
\end{itemize}

\section{Related work}
\subsection{Speaker diarization}
Conventional diarization methods are typically a cascade of four modules: 1) speech activity detection (SAD), 2) speaker embedding extraction, 3) embedding clustering, and 4) overlap handling as an optional process.
Some methods also include an ASR module \cite{india2017lstm,park2019speaker}.
Most studies mainly focus on 2) speech embedding extraction and 3) embedding clustering.
For speaker embeddings, i-vectors \cite{senoussaoui2014study,sell2018diarization}, x-vectors \cite{snyder2018xvectors,diez2019bayesian,xiao2021microsoft}, and d-vectors \cite{wang2018speaker,zhang2019fully} have been explored.
For embedding clustering, earlier works used traditional clustering algorithms, \eg, K-means clustering \cite{shum2013unsupervised,dimitriadis2017developing}, agglomerative hierarchical clustering (AHC) \cite{sell2014speaker,garcia2017speaker,maciejewski2018characterizing}, mean-shift clustering \cite{senoussaoui2014study}, and spectral clustering \cite{wang2018speaker,raj2021multiclass}.
Recently, better clustering methods have been proposed, such as variational Bayes hidden Markov model clustering (VBx) \cite{diez2020analysis,landini2022bayesian}, auto-tuning spectral clustering \cite{park2020auto}, or fully supervised clustering \cite{zhang2019fully,li2021discriminative}.
They are usually used for hard clustering, so most cascaded methods (with some exceptions \cite{huang2020speaker}) cannot deal with speaker overlap.
To make them able to treat speaker overlap, 4) overlap handling should be considered; however, it has sometimes been excluded from methods and evaluations even in very recent studies \cite{sell2018diarization,wang2018speaker,zhang2019fully,li2021discriminative,park2020auto}.
Moreover, 1) speech activity detection has often been ignored in evaluations of cascaded approaches that use oracle speech activities \cite{sell2018diarization,wang2018speaker,zhang2019fully,li2021discriminative,park2020auto}.

Neural-network-based methods that directly produce diarization results from audio are emerging \cite{kinoshita2018listening,von2019all}.
One strength of such methods is that they require no extra modules for SAD or overlap handling.
For some methods, models have been trained for speech separation, and diarization results have been obtained as byproducts \cite{kinoshita2018listening,von2019all}.
Such models have been trained on the basis of clean speech (or time-frequency masks calculated from clean speech); thus, they cannot be trained on real mixtures like DIHARD datasets \cite{ryant2019second,ryant2021third}.
However, EEND-based models are trained to output multiple speakers' speech activities; they do not require clean speech for training and real mixtures can be used.
The original EEND \cite{fujita2019end1,fujita2019end2,fujita2020endtoend} can output diarization results for a fixed number of speakers.
To extend the EEND for an unknown number of speakers, two approaches have been investigated.
One is an attractor-based approach \cite{horiguchi2020endtoend,han2021bwedaeend}, and the other is a speaker-wise conditional EEND (SC-EEND) \cite{fujita2020neural,takashima2021endtoend}.
In this paper, we investigate the attractor-based EEND because it showed better performance compared to SC-EEND.

\subsection{Speech processing based on neural networks for unknown numbers of speakers}
\label{sec:related_work_unknown_num_spk}
While some methods have achieved promising results with a fixed number of output speakers in diarization \cite{fujita2019end1,fujita2019end2,medennikov2020targetspeaker} and speech separation \cite{yu2017permutation,luo2017deep,luo2018tasnet,luo2019conv} contexts, it is challenging to make them able to deal with unknown numbers of speakers.
The difficulty of neural-network-based speech processing for unknown numbers of speakers is that we cannot fix the output dimension.

One possible approach is to determine the maximum number of speakers to decode.
In this case, the number of outputs is set to a sufficiently large value.
Some methods treat a flexible number of speakers by outputting null speech activities if the number of outputs is smaller than the network capacity \cite{luo2018speaker}.
However, this approach did not work well with EEND (see \cite{fujita2020neural}).
In other methods, the number-of-speaker-wise output branches are trained independently, and the most probable is used during inference \cite{zeghidour2021wavesplit}.
In this case, we have to know the maximum number of speakers.
One of the strengths of EEND is that it can be finetuned using a target domain dataset from a pretrained model, but we usually cannot access the maximum number of speakers of the target domain beforehand.
Therefore, a method that does not require that the maximum number of speakers be defined would be preferable.

Another approach is to decode speakers one by one until a stopping condition is satisfied, like SC-EEND \cite{fujita2020neural}.
For speech separation, RSAN \cite{kinoshita2018listening,von2019all} and one-and-rest permutation invariant training (OR-PIT) \cite{takahashi2019recursive} can be used.
The key difference between speech separation and diarization is whether or not the residual output can be defined.
RSAN uses a mask-based approach, in which each time-frequency bin is softly assigned to each speaker so that the process finishes when all the elements of the residual mask become zero.
OR-PIT is time-domain speech separation by which residual output is determined as a mixture that contains other speakers rather than the target speaker.
Both require clean recordings to determine oracle masks or signals.
However, they are not always accessible in the diarization context, in which only multi-talker recordings and speech segments are provided.

In this paper, we adopted an attractor-based approach like deep attractor networks (DANet) \cite{chen2017deep,luo2018speaker}.
While the number of speakers \cite{chen2017deep} or maximum number of speakers \cite{luo2018speaker} is fixed for the original DANet, in this paper, we calculated a flexible number of attractors without defining them.

\subsection{Neural-network-based representative vector calculation}
There have been several efforts to calculate representative vectors from a sequence of embeddings in an end-to-end trainable fashion.
For example, Set Transformer \cite{lee2019set} enables set-to-set transformation, which can be used to calculate cluster centroids from a set of embeddings.
However, the number of outputs has to be known in advance, so it cannot be used for our purpose.
Meier \etal proposed an end-to-end clustering framework \cite{meier2018learning}, in which clustering for all possible number of clusters $K\in\left\{1,\dots,K_\text{max}\right\}$ is performed and the result of the most probable number of clusters is used.
The framework performs the clustering of a flexible number of clusters in an end-to-end manner, but the maximum number of clusters is limited by $K_\text{max}$.
EDA in this paper, in comparison, determines a flexible number of attractors from an input embedding without prior knowledge of the number of speakers.
Thus, we can use datasets of the different maximum number of speakers during pretraining and finetuning.

\section{Method}
In this section, we first introduce the conventional EEND in \autoref{sec:conventional_eend} followed by an explanation of a natural extension of the method called attractor-based EEND in \autoref{sec:proposed}. We also provide novel inference techniques in \autoref{sec:inference}.

\subsection{Conventional end-to-end neural diarization}\label{sec:conventional_eend}
End-to-end neural diarization (EEND) \cite{fujita2019end1,fujita2019end2} is a method for estimating multiple speakers' speech activities simultaneously from an input recording.
Given frame-wise $F$-dimensional acoustic features $\left(\vect{x}_t\right)_{t=1}^T$, where $t\in\left\{1,\dots,T\right\}$ is a frame index, EEND estimates speech activities $\left(\vect{y}_t\right)_{t=1}^T$.
Here, $\vect{y}_t\coloneqq\trans{\left[y_{1,t},\dots,y_{s,t},\dots,y_{S,t}\right]}$ denotes speech activities of $S$ speakers at $t$ defined as
\begin{align}
    y_{s,t}=\begin{cases}
    0&\text{(Speaker $s$ is inactive at $t$)}\\
    1&\text{(Speaker $s$ is active at $t$)}\\
    \end{cases}.
\end{align}
EEND assumes that $y_{s,t}$ is conditionally independent given the acoustic features, namely,
\begin{align}
    P\left(\vect{y}_1,\dots,\vect{y}_T\mid\vect{x}_1,\dots,\vect{x}_T\right)=\prod_{t=1}^{T}\prod_{s=1}^{S}P\left(y_{s,t}\mid\vect{x}_1,\dots,\vect{x}_T\right).
\end{align}
With this assumption, speaker diarization can be regarded as a multi-label classification problem and can thus be easily modeled using a neural network $f_\mathsf{EEND}$ as
\begin{align}
    \left(\vect{p}_1,\dots,\vect{p}_T\right)=f_\mathsf{EEND}\left(\vect{x}_1,\dots,\vect{x}_T\right),
    \label{eq:posteriors}
\end{align}
where $\vect{p}_t\coloneqq\trans{\left[p_{1,t},\dots,p_{S,t}\right]}\in\left(0,1\right)^S$ is the posterior probabilities of $S$ speakers' speech activities at frame index $t$.
The estimation of speech activities $\left(\hat{\vect{y}}_t\right)_{t=1}^{T}$ is 
\begin{align}
    \hat{\vect{y}}_{1},\dots,\hat{\vect{y}}_T&=\argmax_{\vect{y}_1,\dots,\vect{y}_T}P\left(\vect{y}_1,\dots,\vect{y}_T\mid\vect{x}_1,\dots,\vect{x}_T\right),\\
    &=\left(\mathds{1}\left(p_{s,t}>0.5\right)\right)_{\genfrac{}{}{0pt}{2}{1\leq s\leq S}{1\leq t\leq T}},\label{eq:estimation}
\end{align}
where $\mathds{1}\left(\mathrm{cond}\right)$ is an indicator function that returns $1$ if $\mathrm{cond}$ is satisfied and $0$ otherwise.
Note that the threshold value in \autoref{eq:estimation} is always set to 0.5 in this paper for simplicity.

The conventional EEND is implemented as a composition of an embedding part $g: \mathbb{R}^{F\times T}\rightarrow\mathbb{R}^{D\times T}$ and a classification part $h: \mathbb{R}^{D\times T}\rightarrow\left(0,1\right)^{S\times T}$, \ie,
\begin{align}
    f_\mathsf{EEND}=h\circ g.\label{eq:composition}
\end{align}
The first embedding part $g$ converts input acoustic features into $D$-dimensional frame-wise embeddings. 
It is implemented with $N$-stacked encoders, each of which converts a flexible length of embedding sequence $(\vect{e}_t^{(n-1)})_{t=1}^T$ into the same length of embedding sequence $(\vect{e}_t^{(n)})_{t=1}^T$ as
\begin{align}
    \vect{e}_1^{(n)},\dots,\vect{e}_T^{(n)}&=g^{(n)}\left(\vect{e}_1^{(n-1)},\dots,\vect{e}_T^{(n-1)}\right),\\
    \vect{e}_t^{(0)}&=\vect{x}_t\quad(1\leq t\leq T),
\end{align}
where $g^{(n)}$ is the $n$-th encoder layer.
As examples of encoders, bi-directional long short-term memories (BLSTM) \cite{fujita2019end1} and Transformers \cite{fujita2019end2} are exploited in the conventional studies.
In this paper, we used Transformer encoders but without positional encodings to prevent the outputs from being affected by the absolute position of the frames.
Hereafter, for simplicity, we use $\vect{e}_t$ to denote the embeddings from the last encoder, \ie, $\vect{e}_t\coloneqq\vect{e}_t^{(N)}$ for $t\in\left\{1,\dots,T\right\}$.

Then, the classification part $h$ in \autoref{eq:composition} converts the embeddings $\left(\vect{e}_t\right)_{t=1}^T$ to posteriors of speech activities $\left(\vect{p}_t\right)_{t=1}^T$ in \autoref{eq:posteriors}.
It is implemented by using a fully connected layer and an element-wise sigmoid function $\sigma(\cdot)$ that takes a tensor as an argument:
\begin{align}
    \left[\vect{p}_1,\dots,\vect{p}_T\right]&=h(\vect{e}_1,\dots,\vect{e}_T;W_\text{cls},\vect{b}_\text{cls})\\
    &=\sigma\left(\trans{W}_\text{cls}\left[\vect{e}_1,\dots,\vect{e}_T\right]+\vect{b}_\text{cls}\trans{\vect{1}_D}\right)\in\left(0,1\right)^{S\times T},\label{eq:eend_posterior}
\end{align}
where $\trans{\left(\cdot\right)}$ denotes the matrix transpose, $\vect{1}_D$ is $D$-dimensional all-one vector, and $W_\text{cls}\in\mathbb{R}^{D\times S}$ and $\vect{b}_\text{cls}\in\mathbb{R}^S$ are the weight and bias of the fully connected layer, respectively.

EEND outputs posteriors of multiple speakers simultaneously but without any conditions to decide the order of the speakers.
Such a network is optimized by using a permutation-free objective \cite{hershey2016deep,yu2017permutation}, which was originally proposed for multi-talker speech separation.
It computes the loss for all possible speaker assignments between predictions $\left(\vect{p}_t\right)_{t=1}^T$, as introduced in \autoref{eq:posteriors}, and groundtruth labels $\left(\vect{y}_t\right)_{t=1}^T$, and it picks the minimum one for backpropagation as follows.
\begin{align}
    \mathcal{L}_\text{diar}=\frac{1}{TS}\min_{\vect{\phi}\in\Phi\left(S\right)}\sum_{t=1}^T H\left(\vect{y}_t^\phi,\vect{p}_t\right),
    \label{eq:pit_loss}
\end{align}
where $\Phi\left(S\right)$ is a set of all possible permutations of the sequence $\left(1,\dots,S\right)$, $\vect{\phi}\coloneqq\left(\phi_1,\dots,\phi_S\right)$ is the permuted sequence, $\vect{y}_t^\phi\coloneqq\trans{\left[y_{\phi_1,t},\dots,y_{\phi_S,t}\right]}\in\left\{0,1\right\}^S$ is the permuted groundtruth labels using $\vect{\phi}$, and $H\left(\cdot,\cdot\right)$ is the binary cross entropy defined as
\begin{align}
    H\left(\vect{y}_t,\vect{p}_t\right)\coloneqq\sum_{s=1}^{S}\left\{-y_{s,t}\log p_{s,t}-\left(1-y_{s,t}\right)\log\left(1-p_{s,t}\right)\right\}.
    \label{eq:bce}
\end{align}

Compared with cascaded approaches, EEND has two significant strengths.
One is that the cascaded approaches conduct diarization by dividing frame-wise speaker embeddings, so they require SAD as pre-processing and overlap detection and assignment as post-processing.
In contrast, EEND estimates each speaker's speech activities independently, so no extra modules for speech activity detection and overlap detection are needed.
The other strength is that the EEND model can be adapted to the desired domain's dataset, while cascaded approaches typically tune only probabilistic linear discriminant analysis (PLDA) parameters to optimize intra- and inter-speaker similarity between speaker embeddings \cite{sell2014speaker,landini2020but,horiguchi2021hitachi}.

\subsection{Attractor-based end-to-end neural diarization}\label{sec:proposed}
The limitation of the conventional EEND is in the classification part $h$ in \autoref{eq:composition}; the number of output speakers $S$ is fixed by the fully connected layer as in \autoref{eq:eend_posterior}.
One possible way to treat a flexible number of speakers with this fixed-output architecture is to set the number of outputs to be large enough.
However, as discussed in \autoref{sec:related_work_unknown_num_spk}, it requires knowing the maximum number of speakers in advance, and it has been already verified that such a strategy results in poor performance (see \cite{fujita2020neural}).
It is also a problem that the calculation cost of the permutation-free loss increases if we set a large number of speakers to be output.
Therefore, a significant research question is how to output diarization results for a flexible number of speakers.

\begin{figure}[t]
    \centering
    \resizebox{\linewidth}{!}{%
    \includegraphics{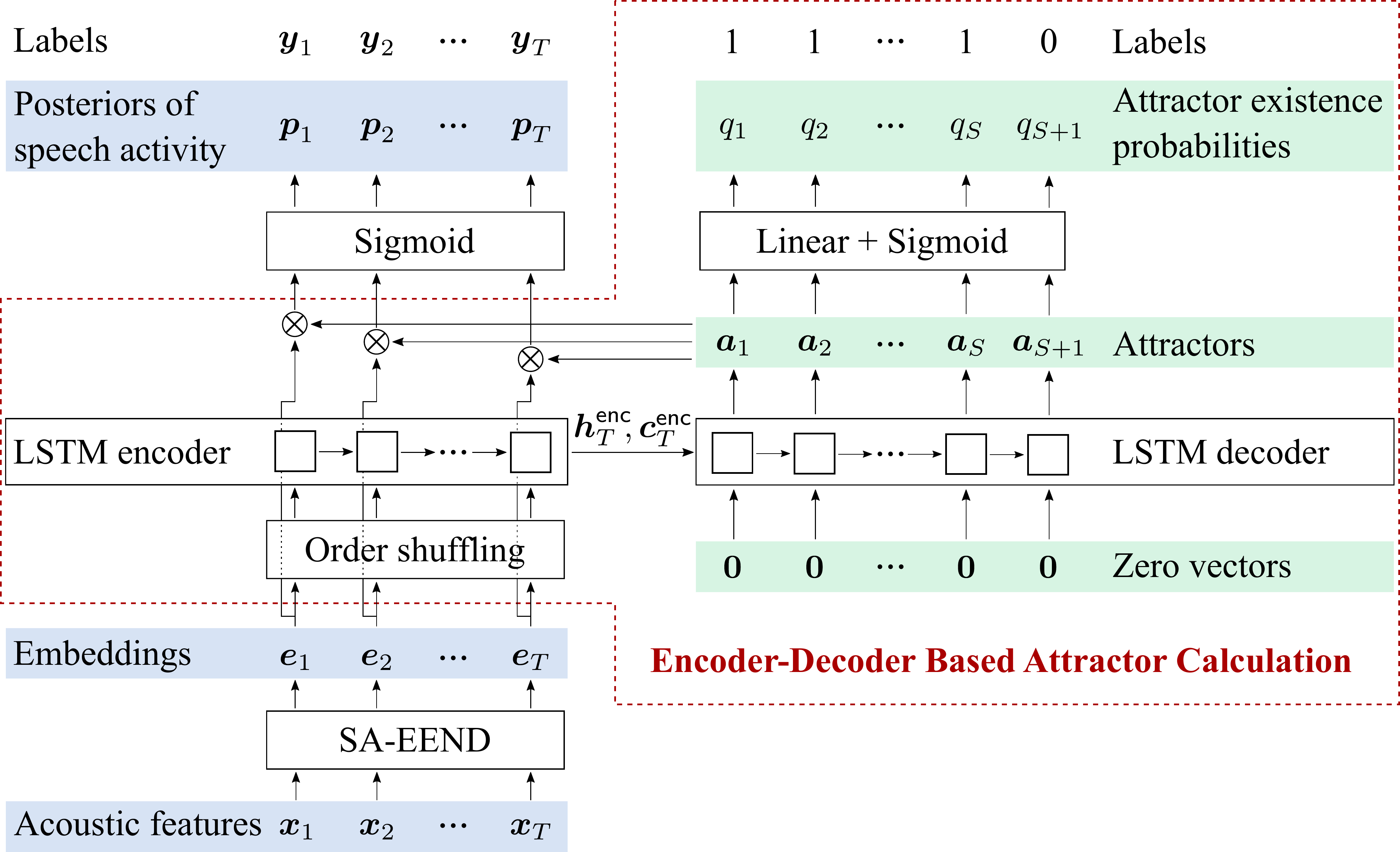}%
    }
    \caption{EEND with encoder-decoder-based attractor calculation (EEND-EDA).}
    \label{fig:overview}
\end{figure}

In this paper, we extend the conventional EEND to handle a flexible number of speakers.
We assume that the embedding part $g$ in \autoref{eq:composition} is implemented in the same manner as the conventional EEND described in \autoref{sec:conventional_eend}.
Given frame-wise $D$-dimensional embeddings $\left\{\vect{e}_t\right\}_{t=1}^T$, our goal is to produce posteriors for a flexible number of speakers in the classification part $h$.
To achieve this goal, we propose a method to calculate a flexible number of speaker-wise attractors from embeddings and then calculate diarization results on the basis of attractors and embeddings.
The proposed method is depicted in \autoref{fig:overview}.

\subsubsection{EDA: Encoder-decoder-based attractor calculation}
EDA converts frame-wise embeddings into speaker-wise attractors using a sequence-to-sequence method with an LSTM encoder-decoder.
The LSTM encoder $h^\mathsf{enc}$ takes the frame-wise embeddings as input and updates its hidden state $\vect{h}_t^\mathsf{enc}$ and cell state $\vect{c}_t^\mathsf{enc}$ as
\begin{align}
    \vect{h}_t^\mathsf{enc}, \vect{c}_t^\mathsf{enc}=h^\mathsf{enc}\left(\vect{e}_{t},\vect{h}_{t-1}^\mathsf{enc},\vect{c}_{t-1}^\mathsf{enc}\right)\qquad\left(t=1,\dots,T\right).
    \label{eq:eda_encoder_chronological}
\end{align}
The hidden and cell states of the encoder are initialized with zero vectors, \ie, $\vect{h}_0^\mathsf{enc}=\vect{c}_0^\mathsf{enc}=\vect{0}$.
The LSTM decoder $h^\mathsf{dec}$ estimates speaker-wise attractors as
\begin{align}
    \vect{h}_s^\mathsf{dec},\vect{c}_s^\mathsf{dec}=h^\mathsf{dec}\left(\vect{0},\vect{h}_{s-1}^\mathsf{dec},\vect{c}_{s-1}^\mathsf{dec}\right)\qquad\left(s=1,2,\dots\right).
    \label{eq:eda_decoder}
\end{align}
We treat the hidden state at each step $\vect{h}_s^\mathsf{dec}\eqqcolon\vect{a}_s\in\left(-1,1\right)^D$ as speaker $s$'s attractor, whose dimensionality $D$ is the same as that of the frame-wise embeddings $\vect{e}_t$.
The hidden and cell states of the decoder are initialized by the final hidden and cell states of the encoder as 
\begin{align}
    \vect{h}_0^\mathsf{dec}&=\vect{h}_T^\mathsf{enc},\\
    \vect{c}_0^\mathsf{dec}&=\vect{c}_T^\mathsf{enc},
\end{align}
which is shown as a right arrow from the LSTM encoder to the LSTM decoder in \autoref{fig:overview}.
In general applications of a sequence-to-sequence method, \eg, speech recognition or machine translation, the output is sentences, \ie, a sequence of words, so the order of output is fixed.
However, EDA cannot determine the order of output speakers in advance because this order is determined by minimizing cross entropy as in \autoref{eq:pit_loss}.
Even if the order could be predetermined, it would not be possible to determine the optimal attractor outputs. 
Thus, the well-known strategy of teacher forcing, for which the optimal outputs with their order have to be known in advance, cannot be used.
Furthermore, the $s$-th attractor can correspond to any speaker that is not contained in the first $(s-1)$ attractors.
To make this attractor calculation procedure fully order-free, we input a zero vector as input at each step as in \autoref{eq:eda_decoder}.
Using zero vectors as inputs provides flexibility to change the number of output speakers across pretraining and finetuning rather than using, for example, trainable parameters.
This is why we chose an LSTM-based encoder-decoder rather than Transformer encoder-decoder, which requires input queries rather than zero vectors.

Here, the input order to the EDA encoder affects the output attractors because EDA is based on a sequence-to-sequence method.
To investigate the effect of the input order, we tried two types of input orders: chronological and shuffled orders.
In the chronological order setting, embeddings are input in the order of frame indexes as in \autoref{eq:eda_encoder_chronological}.
In the shuffled order setting, we use the following instead of \autoref{eq:eda_encoder_chronological} :
\begin{align}
    \vect{h}_t^\mathsf{enc}, \vect{c}_t^\mathsf{enc}=h^\mathsf{enc}\left(\vect{e}_{\psi_t},\vect{h}_{t-1}^\mathsf{enc},\vect{c}_{t-1}^\mathsf{enc}\right)\qquad\left(t=1,\dots,T\right),
\end{align}
where $\left(\psi_1,\dots,\psi_T\right)$ is a randomly chosen permutation of $\left(1,\dots,T\right)$.

The diarization results $\vect{p}_t$ in \autoref{eq:posteriors} are calculated on the basis of the dot product of the frame-wise embeddings and speaker-wise attractors ($\otimes$ in \autoref{fig:overview}):
\begin{align}
    \vect{p}_t=\sigma\left(\trans{A}\vect{e}_t\right)\in\left(0,1\right)^S,\label{eq:eend_eda_posterior}
\end{align}
where $A\coloneqq\left[\vect{a}_1,\dots,\vect{a}_S\right]$ are the speaker-wise attractors.
The posteriors are optimized by using \autoref{eq:pit_loss} in the same manner as the conventional EEND.
This posterior calculation no longer depends on the fully connected layer, which determines the output number of speakers as in \autoref{eq:eend_posterior}; therefore, EDA-based diarization can vary the output number of speakers.

Comparing \autoref{eq:eend_posterior} and \autoref{eq:eend_eda_posterior}, the conventional EEND can also be regarded as using fixed attractors $W_\text{cls}$ (with bias $\vect{b}_\text{cls}$).
In comparison, EDA calculates attractors from an input sequence of embeddings, which makes attractors adaptive to the embeddings.
This makes EEND-EDA more accurate even under the fixed-number-of-speakers condition (see \autoref{tbl:results_twospk}).

\subsubsection{Attractor existence probability}
As in \autoref{eq:eda_decoder}, we can obtain an infinite number of attractors.
To decide when to stop the attractor calculation, we calculate the attractor existence probabilities from the calculated attractors by using a fully connected layer followed by sigmoid activation:
\begin{align}
    q_s=\sigma\left(\trans{\vect{w}_\text{exist}}\vect{a}_s+b_\text{exist}\right),
    \label{eq:attractor_existence_probability}
\end{align}
where $\vect{w}_\text{exist}\in\mathbb{R}^D$ and $b_\text{exist}\in\mathbb{R}$ are trainable weights and bias parameters of the fully connected layer, respectively.

During training, we know the oracle number of speakers $S$, so the training objective of the attractor existence probabilities is based on the first $(S+1)$-th attractors using the binary cross entropy defined in \autoref{eq:bce}:
\begin{align}
    \mathcal{L}_\text{exist}=\frac{1}{S+1}H\left(\vect{l},\vect{q}\right),
    \label{eq:exist_loss}
\end{align}
where
\begin{align}
    \vect{l}&\coloneqq\trans{[\underbrace{1,\dots,1}_{S},0]},\\
    \vect{q}&\coloneqq\trans{\left[q_1,\dots,q_{S+1}\right]}.
\end{align}
The total loss is defined as the weighted sum of $\mathcal{L}_\text{diar}$ in \autoref{eq:pit_loss} and $\mathcal{L}_\text{exist}$ in \autoref{eq:exist_loss} with the weighting parameter $\alpha\in\mathbb{R}_{+}$ as 
\begin{align}
    \mathcal{L}=\mathcal{L}_\text{diar}+\alpha\mathcal{L}_\text{exist}.\label{eq:loss_total}
\end{align}
In this paper, we use $\alpha=1$.
This multi-task loss aims to optimize frame- and speaker-wise posteriors with $\mathcal{L}_\text{diar}$ and attractor existence probabilities with $\mathcal{L}_\text{exist}$.

While \autoref{eq:loss_total} was used for the network optimization in our previous study \cite{horiguchi2020endtoend}, we found that the optimization of $\mathcal{L}_\text{exist}$ inhibits the minimization of $\mathcal{L}_\text{diar}$ during the training of a model with a flexible number of speakers, which is more important for improving diarization accuracy.
Therefore, when a flexible number of speakers' dataset is used for training, we use $\mathcal{L}_\text{exist}$ to update only the fully connected layer parameterized by $\vect{w}_\text{exist}$ and $b_\text{exist}$ in \autoref{eq:attractor_existence_probability}.
This can be implemented by cutting the graph before the fully connected layer to disable backpropagation to the preceding layers.
\begin{comment}
This is performed by using the following instead of \autoref{eq:attractor_existence_probability}:
\begin{align}
    \check{\vect{a}}_s&\leftarrow\mathsf{NoGrad}\left(\vect{a}_s\right),\label{eq:nograd}\\
    q_s&=\frac{1}{1+\exp\left(-\trans{\vect{w}_\text{exist}}\check{\vect{a}}_s+b_\text{exist}\right)},\label{eq:attractor_existence_probability_nograd}
\end{align}
where $\mathsf{NoGrad}\left(\cdot\right)$ denotes the operation for cutting a computational graph to disable back propagation to the preceding layers.
\end{comment}

During inference, we cannot access the oracle number of speakers; thus, it is estimated using $q_s$ in \autoref{eq:attractor_existence_probability} as follows.
\begin{align}
    \hat{S}=\min\left\{s\mid s\in\mathbb{Z}_{+}\wedge q_{s+1}<\tau\right\},
\end{align}
where $\tau\in\left(0,1\right)$ is a thresholding parameter, which is set to $0.5$ in this paper.
We then use the first $\hat{S}$ attractors to calculate posteriors as in \autoref{eq:eend_eda_posterior}.

\subsection{Inference methodology}\label{sec:inference}
\subsubsection{SAD post-processing}\label{sec:sad_postprocessing}
Diarization methods, especially cascaded ones, are sometimes evaluated with oracle speech segments.
When evaluated in such a way, the comparison between cascaded methods and EEND-methods becomes hard, mainly because EEND-based methods perform SAD and diarization simultaneously.
One reason evaluations of cascaded approaches are mainly based on oracle speech segments is to consider speaker errors and SAD errors separately.
It is reasonable to use oracle speech segments to focus on reducing speaker errors.
However, such segments are not accessible in real scenarios, and the existence of SAD errors may worsen the clustering performance, which directly affects the diarization accuracy.
Thus, we believe that SAD errors should also be considered in the context of cascaded methods.
However, it is hard to say how accurate the SAD should be for a fair comparison between cascaded and EEND-based methods.
Therefore, to align with the the cascaded methods, we introduce SAD post-processing for evaluating EEND.
With this method, we can conduct a fair comparison between cascaded and EEND-based methods with the same SAD.
Note that it can be used to improve the diarization performance by eliminating false alarm speech and recovering missed speech when an accurate external SAD system is given.

The SAD post-processing algorithm is described in \autoref{alg:sad_postprocessing}.
Here, we assume that we have SAD results $z_1,\dots,z_T$ in addition to frame- and speaker-wise posteriors $\vect{p}_1,\dots,\vect{p}_T$.
We first estimate speech activities as usual by using \autoref{eq:estimation} (\autoref{algline:sp_decode}).
However, this estimation is not always consistent with SAD results.
Thus, we first filter false alarms (FA) by using SAD results.
For each frame (\autoref{algline:sp_foreach}), if it is estimated that some speakers are active while the speech activity should be zero (\autoref{algline:sp_filter1}), we update the estimations with a zero vector (\autoref{algline:sp_filter2}).
This procedure will always improve DER if $z_1,\dots,z_T$ are the oracle speech activities.
We also recover missed frames (MI) if no speaker is estimated as active while the speech activity is one (\autoref{algline:sp_recover1}).
For each of such frames, we treat the speaker with the highest posterior as an active speaker (\autoref{algline:sp_recover2}--\autoref{algline:sp_recover3}).
Including the oracle SAD as input will also improve the DER because missed-frame errors are replaced by correct estimation or at least speaker errors.

\begin{algorithm}[t]
    \SetAlgoLined
	\DontPrintSemicolon
	\caption{SAD post-processing.}
	\label{alg:sad_postprocessing}
	\SetAlgoVlined
	\SetKwInOut{Input}{Input}\SetKwInOut{Output}{Output}
	\SetKw{In}{in}
	\SetKw{Continue}{continue}
	\SetKw{And}{and}
	\Input{$\left(\vect{p}_1,\dots,\vect{p}_T\right)\in\left(0,1\right)^{S\times T}$\tcp*{Frame-wise posteriors}}
	\myinput{$\left(z_1,\dots,z_T\right)\in\left\{0,1\right\}^T$\tcp*{SAD results}}
	\Output{$\left(\hat{\vect{y}_1},\dots,\hat{\vect{y}_T}\right)\in\left\{0,1\right\}^{S\times T}$ \tcp*{Speech activities}}
	\BlankLine
	Compute $\hat{\vect{y}_1},\dots,\hat{\vect{y}_T}$ using \autoref{eq:estimation}\tcp*{Initial results}\label{algline:sp_decode}
	\ForEach{$t\in\{1,\dots,T\}$}{\label{algline:sp_foreach}
	    \uIf(\tcp*[f]{Filter FA}){$\norm{\hat{\vect{y}_t}}_1>0\land z_t=0$}{\label{algline:sp_filter1}
	        $\hat{\vect{y}_t}\leftarrow\trans{\left[0,\dots,0\right]}$\;\label{algline:sp_filter2}
	    }
	    \ElseIf(\tcp*[f]{Recover MI}){$\norm{\hat{\vect{y}_t}}_1=0\land z_t=1$}{\label{algline:sp_recover1}
	        $s^*\leftarrow\argmax_{s\in 1,\dots,S}{\vect{p}_t}$\;\label{algline:sp_recover2}
	        $\hat{\vect{y}_t}\leftarrow\trans{[0,\dots,0,\underset{s^*}{\underset{\scriptscriptstyle\wedge}{1}},0,\dots,0]}\in\left\{0,1\right\}^S$\;\label{algline:sp_recover3}
	    }
	}
\end{algorithm}

\subsubsection{Iterative inference}\label{sec:iterative_inference}
Even if the model is trained to output a flexible number of speakers, the output number of speakers is empirically limited by the maximum number of speakers in a recording observed during pre-training (see \autoref{tbl:ablation_study}).
How to output the results of more than $N$ speakers even if the model is trained on at most $N$-speaker mixtures is still an open question.
In this paper, we propose an iterative inference method to produce results for more than $N$ speakers by applying EEND decoding with iterative frame selection.

Preliminarily, we first reveal the characteristics of the EEND models that consist of stacked Transformer encoders and EDA. 
A Transformer encoder involves neither recurrence nor convolutional calculation, and we do not use positional encoding in this paper; thus, the embedding part $g$ in \autoref{eq:composition} is an order-free transformation.
EDA contains an LSTM encoder-decoder, but if the order of the input sequence to EDA is shuffled, we can say that EDA does not depend on the input order, so the EDA's classification part $h$ in \autoref{eq:composition} is also an order-free function.
Therefore, EEND-EDA does not depend on the order of the input features, which makes it possible to process features that are not extracted at equal intervals along the time axis, as in EEND as post-processing \cite{horiguchi2021endtoend}.
The proposed iterative inference also utilizes this characteristic.

\begin{algorithm}[t]
    \SetAlgoLined
	\DontPrintSemicolon
	\caption{Iterative inference.}
	\label{alg:iterative_inference}
	\SetAlgoVlined
	\SetKwInOut{Input}{Input}\SetKwInOut{Output}{Output}
	\SetKw{In}{in}
	\SetKw{Continue}{continue}
	\SetKw{Break}{break}
	\Input{$\vect{x}_1,\dots,\vect{x}_T$\tcp*{Acoustic features}}
	\myinput{$f_\mathsf{EEND}$\tcp*{EEND model}}
	\myinput{$S_\text{\rm max}\in\mathbb{N}$ \tcp*{Max \#Speakers that EEND can output}}
	\Output{$\hat{Y}\in\left\{0,1\right\}^{S\times T}$}
	\BlankLine
	$\mathcal{T}\leftarrow\left\{1,\dots,T\right\}$\tcp*{Frame set}\label{algline:ii_init_t}
	%$n=1$\tcp*{Iteration counter}\label{algline:ii_init_counter}
	%\While{\rm true}{\label{algline:ii_while}
	\For{$n\leftarrow 1$~\KwTo~$\infty$}{
    	Compute $\hat{Y}^{(n)}$ by \autoref{eq:decode_selected_frames}, \autoref{eq:decode_unselected_frames}, and \autoref{eq:estimation}\tcp*{Decoding}\label{algline:ii_decoding}
    	Update $\mathcal{T}$ by \autoref{eq:silence_frame_selection}\tcp*{Silence frame selection}\label{algline:ii_silence_frame_selection}
    	\If{$S^{(n)}<S_\text{\rm max}\lor\abs{\mathcal{T}}=0$}{\label{algline:ii_break1}
    	    \Break\;\label{algline:ii_break2}
    	}
    	%$n\leftarrow n+1$\;\label{algline:ii_update_count}
	}
	$\hat{Y}\leftarrow\begin{bmatrix}\hat{Y}^{(1)}\\\vdots\\\hat{Y}^{(n)}\end{bmatrix}$\label{algline:ii_concat}
\end{algorithm}

\begin{figure}[t]
    \centering
    \includegraphics[width=\linewidth]{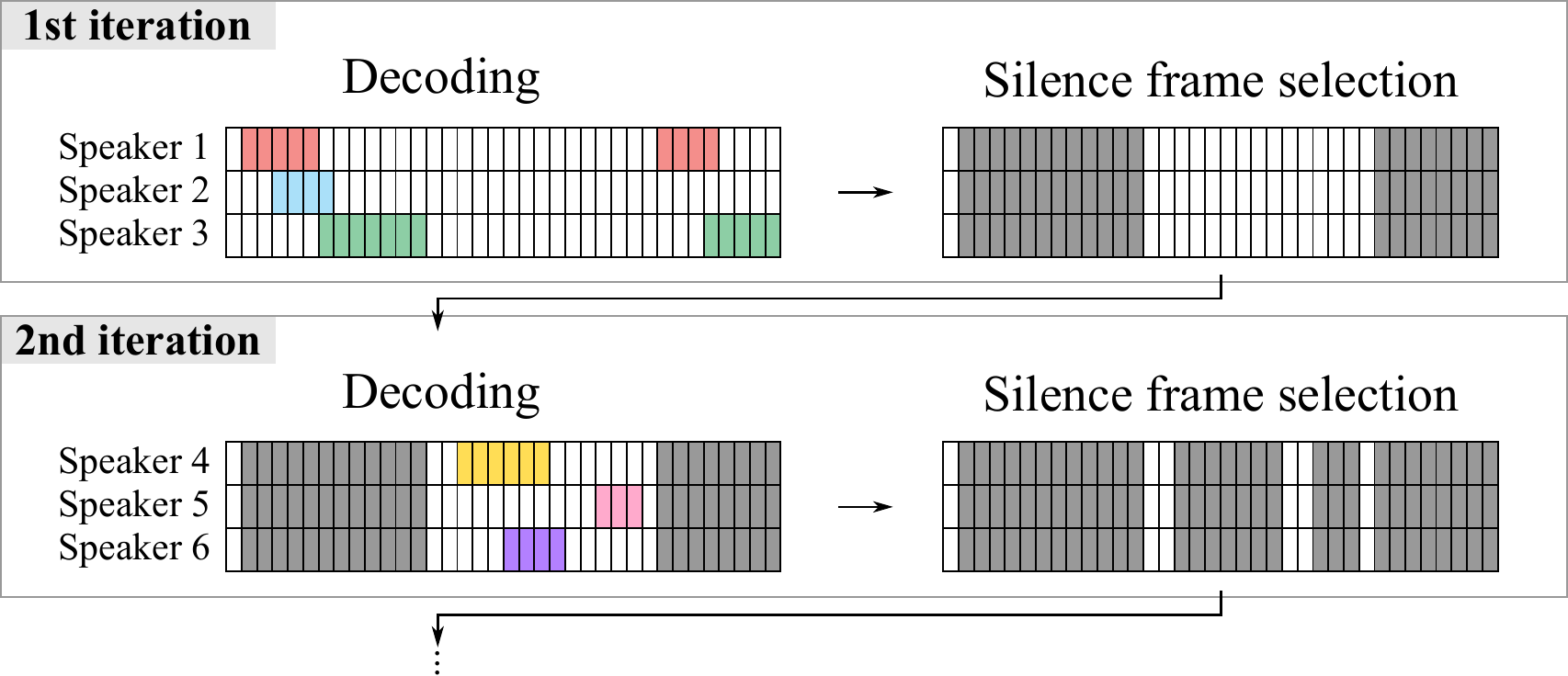}
    \caption{Iterative inference in the case of $S_\text{max}=3$.}
    \label{fig:iterative_inference}
\end{figure}

\autoref{alg:iterative_inference} shows the algorithm of iterative inference. 
In the algorithm, two processes are iteratively conducted: decoding and silence frame selection.
Each process at the $n$-th iteration is described as follows.
\begin{enumerate}
    \item \textbf{Decoding} (\autoref{algline:ii_decoding}): Acoustic features $\vect{x}_t$ of the selected frames $\mathcal{T}$ are fed into EEND, and the corresponding posteriors $\vect{p}_t^{(n)}\in\left(0,1\right)^{S^{(n)}}$ are obtained as
    \begin{align}
        \left(\vect{p}_t^{(n)}\right)_{t\in\mathcal{T}}\leftarrow f_\mathsf{EEND}\left(\left(\vect{x}_t\right)_{t\in\mathcal{T}}\right),\label{eq:decode_selected_frames}
    \end{align}
    where $S^{(n)}\in\left\{0,\dots,S_\text{max}\right\}$ is the number of decoded speakers.
    The posteriors of the frames that are not in $\mathcal{T}$ are set to zero as
    \begin{align}
        \vect{p}_t^{(n)}\leftarrow[\underbrace{0,\dots,0}_{S^{(n)}}\trans{]}\quad\left(t\in\{1,\dots,T\}\setminus\mathcal{T}\right).\label{eq:decode_unselected_frames}
    \end{align}
    With the posteriors $\vect{p}_t^{(n)}$ for $t\in\left\{1,\dots,T\right\}$, diarization results $\hat{Y}^{(n)}=\left(\hat{\vect{y}}_1^{(n)},\dots,\hat{\vect{y}}_T^{(n)}\right)$ are computed using \autoref{eq:estimation}.
    Note that $\hat{Y}^{(n)}$ corresponds to the speech activities of the $((n-1)S_\text{max}+1)$-th through $((n-1)S_\text{max}+S^{(n)})$-th speakers.
    %\item \textbf{Silence frame selection}: Given the diarization results decoded at the $n$-th iteration, we update the frame set to contain the remaining silent frames as
    \item \textbf{Silence frame selection} (\autoref{algline:ii_silence_frame_selection}): Given the diarization results decoded at the $n$-th iteration, we select the frames in which no speaker is active to update $\mathcal{T}$ as
    \begin{align}
        \mathcal{T}\leftarrow\left\{t\relmid t\in\mathcal{T},\norm{\hat{\vect{y}}_t^{(n)}}_1=0\right\}.\label{eq:silence_frame_selection}
    \end{align}
\end{enumerate}
The above processes start with the initial value of $\mathcal{T}$ as the set of all frames $\left\{1,\dots,T\right\}$ (\autoref{algline:ii_init_t}), and last until $\mathcal{T}$ becomes the empty set or when it is assumed that all the speakers are decoded (\autoref{algline:ii_break1}--\autoref{algline:ii_break2}).
Here, we assume that all the speakers are decoded if the number of output speakers $S^{(n)}$ is smaller than the maximum output of EEND $S_\text{max}$.

After the iterative process is finished, the final results $\hat{Y}$ are obtained by concatenating the results calculated at each iteration (\autoref{algline:ii_concat}).
With iterative inference, the number of speakers to be decoded is no longer limited by the training dataset.
The iterative inference workflow when $S_\text{max}=3$ is also illustrated in \autoref{fig:iterative_inference}.

\subsubsection{Iterative inference with DOVER-Lap (or iterative inference+)}\label{sec:iterative_inference_plus}
Despite iterative inference being able to produce more than $S_\text{max}$ speakers' speech activities, it has a potential problem in that the speech activities of two speakers decoded at different iterations never overlap.
For example, the $(S_\text{max}+1)$-th speaker's speech activities never overlap with those of the first $S_\text{max}$ speakers.
This is because the frames in which the first $S_\text{max}$ speakers are active will not be processed in the second iteration.
To ease this problem, we introduce DOVER-Lap \cite{raj2021doverlap}, which is the extension of DOVER \cite{stolcke2019dover}.
Both of them are methods for combining multiple diarization results on the basis of majority voting, but unlike DOVER, DOVER-Lap take speaker overlap into account.
We used a modified version of DOVER-Lap presented in \cite{horiguchi2021hitachi}, in which the speaker assignment strategy when multiple speakers were ranked equally was slightly different from the original DOVER-Lap \cite{raj2021doverlap}.
Note that we did not use a hypothesis-wise weighting of DOVER-Lap, which is also introduced in \cite{horiguchi2021hitachi}.

\begin{algorithm}[t]
    \SetAlgoLined
	\DontPrintSemicolon
	\caption{Iterative inference with DOVER-Lap (or iterative inference+).}
	\label{alg:iterative_inference_plus}
	\SetAlgoVlined
	\SetKwInOut{Input}{Input}\SetKwInOut{Output}{Output}
	\SetKw{In}{in}
	\SetKw{Continue}{continue}
	\SetKw{Break}{break}
	\Input{$\vect{x}_1,\dots,\vect{x}_T$\tcp*{Acoustic features}}
	\myinput{$f_\mathsf{EEND}$\tcp*{EEND model}}
	\myinput{$S_\text{\rm max}\in\mathbb{N}$ \tcp*{Max \#Speakers that EEND can output}}
	\Output{$\hat{Y}\in\left\{0,1\right\}^{S\times T}$}
	\BlankLine
	\For{$S_\text{\rm limit}=1$~\KwTo~$S_\text{\rm max}$}{\label{algline:iip_vary_s_limit}
		$\mathcal{T}\leftarrow\left\{1,\dots,T\right\}$\tcp*{Frame set}\label{algline:iip_init_t}
    	%$n=1$\tcp*{Iteration counter}\label{algline:iip_init_counter}
    	%\While{\rm true}{
	    \For{$n\leftarrow 1$~\KwTo~$\infty$}{
        	Compute $\hat{Y}^{(n)}$ by \autoref{eq:decode_selected_frames}, \autoref{eq:decode_unselected_frames}, \autoref{eq:estimation}\tcp*{Decoding}
        	\If{$n=1$}{\label{algline:iip_limit1}
            	Limit the number of speakers in $\hat{Y}^{(n)}$ by \autoref{eq:limit_n_spks}\;\label{algline:iip_limit2}
            }
        	Update $\mathcal{T}$ by \autoref{eq:silence_frame_selection}\tcp*{Silence frame selection}
        	\If{$S^{(n)}<S_\text{\rm max}\lor\abs{\mathcal{T}}=0$}{
        	    \Break\;\label{algline:iip_break}
        	}
        	%$n\leftarrow n+1$\;\label{algline:iip_update_count}
    	}
    	$\hat{Y}_{S_\text{limit}}\leftarrow\begin{bmatrix}\hat{Y}^{(1)}\\\vdots\\\hat{Y}^{(n)}\end{bmatrix}$\label{algline:iip_concat}
	}
	$\hat{Y}\leftarrow\text{DOVER-Lap}\left(\hat{Y}_1,\dots,\hat{Y}_{S_\text{\rm max}}\right)$\;\label{algline:iip_doverlap}
\end{algorithm}

The algorithm of iterative inference incorporated with DOVER-Lap is shown in \autoref{alg:iterative_inference_plus}.
In this paper, we refer to this inference as iterative inference+.
The difference from the iterative inference in \autoref{alg:iterative_inference} is that we limit the number of speakers to decode at the first iteration with $S_\text{limit}(\leq S_\text{max})$ (\autoref{algline:iip_limit1}--\autoref{algline:iip_limit2}).
After the decoding step at the first iteration using \autoref{eq:decode_selected_frames}, \autoref{eq:decode_unselected_frames}, and \autoref{eq:estimation}, we choose at most the first $S_\text{limit}$ speakers' speech activities from $\hat{Y}^{(1)}\coloneqq\left(\hat{y}_{s,t}\right)_{s,t}$ as
\begin{align}
    \hat{Y}^{(1)}\leftarrow\left(\hat{y}_{s,t}\right)_{\genfrac{}{}{0pt}{2}{1\leq s\leq \min\left(S^{(1)}, S_\text{limit}\right)}{1\leq t\leq T}}.\label{eq:limit_n_spks}
\end{align}
The other procedures are the same as those in \autoref{alg:iterative_inference}, and finally, we obtain $S_\text{limit}$-wise diarization results $Y_{S_\text{limit}}$ (\autoref{algline:iip_concat}).

In iterative inference+, $S_\text{limit}$ is varied from 1 to $S_\text{max}$ (\autoref{algline:iip_vary_s_limit}), which results in $S_\text{max}$ diarization results for each recording.
We then combine them by using DOVER-Lap to obtain the final result $\hat{Y}$ (\autoref{algline:iip_doverlap}).
With this procedure, the $k$-th speaker's speech activities can be overlapped with those of the $\max\left(1,\left(k-S_\text{max}+1\right)\right)$-th to $\left(k+S_\text{max}-1\right)$-th speakers.

\section{Experiments}
\subsection{Datasets}
\subsubsection{Simulated datasets}
To train the EEND-EDA model, we created simulated speech mixtures from single-speaker recordings of the following corpora.
\begin{itemize}
    \item Switchboard-2 (Phase I \& II \& III)
    \item Switchboard Cellular (Part 1 \& 2)
    \item NIST Speaker Recognition Evaluation (2004 \& 2005 \& 2006 \& 2008)
\end{itemize}
Note that these corpora are compatible with the Kaldi CALLHOME x-vector recipe\footnote{\label{footnote:kaldi_callhome}\url{https://github.com/kaldi-asr/kaldi/tree/master/egs/callhome_diarization/v2}}.

We used the following simulation protocol to create multi-talker mixtures from single-speaker recordings:
\begin{enumerate}
    \item Select $N$ speakers,
    \item For each speaker, randomly sample speech segments and concatenate them with silences that are interlaid between speech segments,
    \item For each of the $N$ long recordings created, randomly select a room impulse response and convolve it with the recording,
    \item Mix the $N$ long recordings and a noise signal with a randomly determined signal-to-noise ratio.
\end{enumerate}
The detailed algorithm for creating simulated mixtures can be found in \cite{fujita2019end1}.
In the second process, we assume that the occurrence of an utterance is a Poisson process, so the duration of the silence between speech segments follows the exponential distribution $\frac{1}{\beta}\exp\left(-\frac{x}{\beta}\right)$, where $\beta$ is the mean value.
$\beta$ can be used to control the overlap ratio of the mixtures.
To obtain a similar overlap ratio among various numbers of speakers, we varied $\beta$ according to the number of speakers as summarized in \autoref{tbl:dataset_sim}.

\begin{table}[t]
    \centering
    \caption{Datasets of simulated mixtures.}
    \label{tbl:dataset_sim}
    \begin{tabular}{@{}lccccc@{}}
        \toprule
        Dataset&Split&\#Spk&\#Mixtures&$\beta$&Overlap ratio (\%)\\\midrule
        Sim1spk &Train& 1 & 100,000&2&0.0\\
        &Test& 1 & 100,000&2&0.0\\\midrule
        Sim2spk &Train& 2&100,000&2&34.1\\
        &Test& 2&500&2&34.4\\
        &Test& 2&500&3&27.3\\
        &Test& 2&500&5&19.1\\\midrule
        Sim3spk &Train& 3&100,000&5&34.2\\
        &Test& 3&500&5&34.7\\
        &Test& 3&500&7&27.4\\
        &Test& 3&500&11&19.2\\\midrule
        Sim4spk &Train& 4&100,000&9&31.5\\
        &Test& 4&500&9&32.0\\\midrule
        Sim5spk &Train& 5&100,000&13&30.3\\
        &Test& 5&500&13&30.7\\
        \bottomrule
    \end{tabular}
\end{table}

\subsubsection{Real datasets}
For real datasets, we employed five multi-talker datasets below.
\begin{itemize}
    \item \textbf{CALLHOME} \cite{callhome}: A dataset that consists of telephone conversations whose average duration is two minutes. We used the splits provided in the Kaldi x-vector recipe\textsuperscript{\ref{footnote:kaldi_callhome}}, which are denoted as Part 1 and Part 2, respectively. Two- and three-speaker subsets were used in the fixed-number-of-speakers evaluations, which are denoted as CALLHOME-2spk and CALLHOME-3spk.
    \item \textbf{CSJ} \cite{maekawa2003corpus}: A dataset that consists of monologues and dialogues of Japanese speech. In this paper, we used the dialogue part of the dataset. The average duration of the recordings is about 13 minutes. Following \cite{kanda2019guided}, we used 54 dialogue recordings out of 58.
    \item \textbf{AMI headset mix} \cite{carletta2007unleashing}: A meeting dataset that consists of 100 hours of multi-modal meeting recordings. Each meeting session is about 30 minutes. We used \textit{headset mix} recordings, which were obtained by mixing the headset recordings of all the participants. We used the split and reference RTTMs provided in the VBx paper \cite{landini2022bayesian}.
    \item \textbf{DIHARD II} \cite{ryant2019second}: A dataset used in the second DIHARD challenge. We used single-channel audio, which is used for tracks 1 and 2. The dataset consists of recordings from 11 domains (including telephone data) with an average duration of about 7 minutes.
    \item \textbf{DIHARD III} \cite{ryant2021third}: A dataset used in the third DIHARD challenge. It also consists of recordings from 11 domains (including telephone data) with an average duration of about 8 minutes. The test set has two evaluation conditions called \textit{core} and \textit{full}. The core set is a subset of the full set, in which the recordings are selected to balance the duration of each domain. In terms of the number of speakers, the full set contains more recordings of two speakers than the core set.
\end{itemize}
Their statistics are summarized in \autoref{tbl:dataset_real}.
Note that the recordings in CSJ, AMI, DIHARD II, and DIHARD III were sampled at \SI{16}{kHz}, so we downsampled them to \SI{8}{kHz} to be aligned with those of the simulated datasets.
We also note that the recordings of the CSJ corpus are in stereo, so we mixed them to create monaural recordings.

\begin{table}[t]
    \centering
    \caption{Datasets of real recordings.}
    \label{tbl:dataset_real}
    \resizebox{\linewidth}{!}{%
    \begin{tabular}{@{}lcccc@{}}
        \toprule
        Dataset & Split &\#Spk&\#Mixtures & Overlap ratio (\%)\\\midrule
        CALLHOME-2spk \cite{callhome}& Part 1 & 2 & 155&14.0\\
        &Part 2 & 2 & 148&13.1\\\midrule
        CSJ \cite{maekawa2003corpus}& --- & 2 & 54&20.1\\\midrule
        CALLHOME-3spk \cite{callhome}&Part 1&3&61&19.6\\
        &Part 2&3&74&17.0\\\midrule
        CALLHOME \cite{callhome}& Part 1 &2--7&249&17.0\\
         & Part 2 &2--6&250&16.7\\\midrule
        AMI headset mix \cite{carletta2007unleashing}&Train&3--5&136&13.4\\
        &Dev&4&18&14.1\\
        &Test&3--4&16&14.6\\\midrule
        DIHARD II \cite{ryant2019second}&Dev&1--10&192&9.8\\
        &Test&1--9&194&8.9\\\midrule
        DIHARD III \cite{ryant2021third}&Dev&1--10&254&10.7\\
        &Test (Core)& 1--9&184&8.8\\
        &Test (Full)& 1--9&259&9.2\\\bottomrule
    \end{tabular}%
    }
\end{table}

\subsection{Training}
For the embedding part $g$ in \autoref{eq:composition} of the proposed EEND-EDA, we used four-stacked Transformer encoders with four attention heads without positional encodings, each of which outputs 256-dimensional frame-wise embeddings.
The inputs for the model were log-scaled Mel-filterbank-based features.
We first extracted 23-dimensional log-scaled Mel-filterbanks with a frame length of \SI{25}{\ms} and frame shift of \SI{10}{\ms}.
Each of them was then concatenated with those of the preceding and following seven frames, followed by subsampling with a factor of 10.
As a result, a $345~(=23\times15)$ dimensional acoustic feature was extracted for each \SI{100}{\ms}.

In this paper, we evaluated EEND-EDA for both fixed-numbers-of-speakers and unknown-numbers-of-speakers conditions; thus, a model was trained for each purpose.
For the fixed-number-of-speakers evaluation, the model was first trained on the Sim$k$spk training set for 100 epochs and evaluated on the Sim$k$spk test set.
We also adapted the model to CALLHOME-$k$spk for another 100 epochs to evaluate the model on real recordings.
We used $k\in\{2,3\}$ in this paper.
For the unknown-number-of-speakers evaluation, the model that was trained on Sim2spk was finetuned by using the concatenation of Sim\{1,2,3,4\}spk or Sim\{1,2,3,4,5\}spk for 50 epochs.
The model was also adapted to each target dataset for another 500 epochs.

For network training using simulated mixtures, we used the Adam optimizer \cite{kingma2015adam} with the Noam scheduler \cite{vaswani2017attention} with 100,000 warm-up steps.
For adaptation, we also used the Adam optimizer but with a fixed learning rate of $1\times10^{-5}$.
For efficient batch processing during training, we split each recording into 500 frames when using Sim$k$spk and 2000 frames when using the adaptation sets.
The batch size for training was set to 64.
Note that an entire recording is fed into the network without splitting during inference.

\subsection{Evaluation}
As an evaluation metric, we used diarization error rates (DERs) defined as
\begin{align}
    \mathrm{DER}=\frac{T_\mathrm{MI}+T_\mathrm{FA}+T_\mathrm{CF}}{T_\mathrm{Speech}},
\end{align}
where $T_\mathrm{Speech}$, $T_\mathrm{MI}$, $T_\mathrm{FA}$, and $T_\mathrm{CF}$ denote the duration of total speech, missed speech, false alarm speech, and speaker confusion, respectively.
Following the prior work in \cite{fujita2019end1,kanda2019simultaneous}, we used \SI{0.25}{\second} of collar tolerance at each speech boundary for the Sim$k$spk, CALLHOME, and CSJ evaluation.
For AMI, DIHARD II, and DIHARD III, we allowed no collar tolerance and used a subsampling factor of 5 during inference, which results in acoustic features extracted every \SI{50}{\ms}, to obtain more fine-grained results.
We emphasize that speaker overlaps were NOT excluded from the evaluations.

We also report Jaccard error rates (JERs) in addition to DERs.
To calculate JER, first, the optimal assignment between reference and system speakers is calculated.
JER is the average score of each reference speaker defined as
\begin{align}
    \mathrm{JER}=\frac{1}{S_\text{ref}}\sum_{s=1}^{S_\text{ref}}\frac{T_\mathrm{FA}^{(s)}+T_\mathrm{MI}^{(s)}}{T_\mathrm{Union}^{(s)}},
\end{align}
where $S_\text{ref}$ is the number of reference speakers, and $T_\mathrm{MI}^{(s)}$ and $T_\mathrm{FA}^{(s)}$ are the duration of the missed and false alarm speech calculated between speech activities of the $s$-th reference speaker and the paired system speaker, respectively.
$T_\mathrm{Union}^{(s)}$ is the time duration in which at least one of the $s$-th reference speakers of a paired system speaker is active.

\section{Results}
\subsection{Fixed numbers of speakers}
\subsubsection{Two-speaker experiment}
First, we evaluated our method under the two-speaker condition.
In this case, the model was first trained on Sim2spk and then adapted to CALLHOME-2spk Part 1.
For the EEND-based methods, we used the model trained on Sim2spk to evaluate the simulated datasets and the one adapted to CALLHOME-2spk Part 1 to evaluate CALLHOME-2spk Part 2 and CSJ.
For EEND-EDA, we used the first two output attractors for speech activity calculation.

\autoref{tbl:results_twospk} shows the results of the two-speaker evaluation.
We observed that the proposed method with the shuffled order setting achieved the best DERs.
Despite EEND-EDA being designed to deal with flexible numbers of speakers, it outperformed the conventional EENDs, \ie, BLSTM-EEND and SA-EEND, which output diarization results for fixed numbers of speakers.
This is because the conventional EEND can be regarded as a fixed-attractor-based method, while EEND-EDA is an adaptive-attractor-based method as described in the last paragraph of \autoref{sec:proposed}.
This flexibility of attractors makes the proposed method more accurate even in fixed-number-of-speakers evaluations.
In terms of the order of the input to EDA, shuffled sequences always performed better than chronologically ordered sequences.
It indicates that the global context is more important than the temporal context to calculate attractors.

\begin{table}[t]
    \centering
    \caption{DERs (\%) for two-speaker evaluations. \SI{0.25}{\second} of collar tolerance was allowed.}
    \label{tbl:results_twospk}
    \resizebox{\linewidth}{!}{%
    \begin{tabular}{@{}lccccc@{}}
        \toprule
        &\multicolumn{3}{c}{Simulated}&\multicolumn{2}{c}{Real}\\\cmidrule(lr){2-4}\cmidrule(l){5-6}
        Method&$\beta=2$&$\beta=3$&$\beta=5$&CALLHOME-2spk&CSJ\\\midrule
        i-vector + AHC & 33.74 & 30.93 & 25.96 & 12.10 & 27.99\\
        x-vector (TDNN) + AHC & 28.77 & 24.46 & 19.78 & 11.53 & 22.96\\
        BLSTM-EEND \cite{fujita2019end1} & 12.28 & 14.36 & 19.69 & 26.03 & 39.33\\
        SA-EEND \cite{fujita2019end2} &4.56&4.50&3.85&9.54&20.48\\
        EEND-EDA (Chronol.)&3.07&2.74&3.04&8.24&18.89\\
        EEND-EDA (Shuffled)&\textbf{2.69}&\textbf{2.44}&\textbf{2.60}&\textbf{8.07}&\textbf{16.27}\\
        \bottomrule
    \end{tabular}%
    }
\end{table}

\subsubsection{Three-speaker experiment}
We also evaluated the method under the three-speaker condition.
We first trained the model on Sim3spk and then adapted it to CALLHOME-3spk Part 1.
We validated the performance on Sim3spk using the model trained on Sim3spk and that on CALLHOME-3spk Part 2 using the model adapted to CALLHOME-3spk Part 1.
We used the first three attractors to evaluate EEND-EDA's performance.
As shown in \autoref{tbl:results_threespk}, EEND-EDA with sequence shuffling performed best on both simulated and real datasets.

\begin{table}[t]
    \centering
    \caption{DERs (\%) for three-speaker evaluations. \SI{0.25}{\second} of collar tolerance was allowed.}
    \label{tbl:results_threespk}
    \resizebox{\linewidth}{!}{%
    \begin{tabular}{@{}lccccc@{}}
        \toprule
        &\multicolumn{3}{c}{Simulated}&Real\\\cmidrule(lr){2-4}\cmidrule(l){5-5}
        Method&$\beta=2$&$\beta=3$&$\beta=5$&CALLHOME-3spk\\\midrule
        x-vector (TDNN) + AHC & 31.78 & 26.06 & 19.55 & 19.01\\
        SA-EEND \cite{fujita2019end2} &8.69 & 7.64 & 6.92 & 14.00\\
        EEND-EDA (Chronol.)&13.02&11.65&10.41&15.86\\
        EEND-EDA (Shuffled)&\textbf{8.38}&\textbf{7.06}&\textbf{6.21}&\textbf{13.92}\\
        \bottomrule
    \end{tabular}%
    }
\end{table}

\subsubsection{Effect of input order}
\begin{table*}[t]
    \centering
    \caption{DERs for Sim2spk (overlap ratio: \SI{34.4}{\percent}) using various types of sequences.}
    \label{tbl:various_sequence}
    \resizebox{\linewidth}{!}{%
    \begin{tabular}{@{}lcccccccccccc@{}}
        \toprule
        &\multicolumn{2}{c}{Using whole sequence}&\multicolumn{5}{c}{Subsample $1/N$}&\multicolumn{5}{c}{Using the last $1/N$}\\\cmidrule(lr){2-3}\cmidrule(lr){4-8}\cmidrule(l){9-13}
        Method&Chronol. & Shuffled & $N=2$ & $N=4$ & $N=8$ & $N=16$ & $N=32$ & $N=2$ & $N=4$ & $N=8$& $N=16$ & $N=32$\\\midrule
        EEND-EDA (Train: Chronol.)&3.07&30.04&3.54&7.32&14.48&21.13&27.18&3.67&4.97&5.40&6.11&7.68\\
        EEND-EDA (Train: Shuffled)&2.69&2.69&2.70&2.68&2.79&3.09&5.08&3.36&5.92&7.46&8.59&10.65\\
        \bottomrule
    \end{tabular}%
    }
\end{table*}

For a better understanding of EDA, we tried various types of sequences as inputs to the models, each of which was trained on chronologically ordered sequences and shuffled sequences.
We evaluated matched and unmatched conditions of orders, and we also evaluated the effect of reducing the sequence length by subsampling or using the last $1/N$ part of the sequences.
\autoref{tbl:various_sequence} shows the results on Sim2spk ($\beta=2$).
The EEND-EDA that was trained on chronologically ordered sequences performed well on chronologically ordered sequences but did poorly on shuffled sequences.
It was also affected by subsampling, while it was slightly influenced by using the last $1/N$ part.
These results indicate that the length of each utterance is an important factor to decide the output attractors for the model trained on chronologically ordered sequences.
On the other hand, when the model was trained on shuffled sequences, it was not that affected by the order of sequences nor subsampling.
However, when the last $1/N$ of the sequences were used, its performance degradation was worse than the model trained on chronologically ordered sequences.
These results indicate that EDA trained on shuffled sequences captured the distribution of embeddings; thus, subsampling did not affect the performance that much, while using the last $1/N$, \ie, biased sampling, degraded the DERs.

\subsubsection{Embedding visualization}
\begin{figure}[t]
    \centering
    \subfloat[][Conventional EEND \cite{fujita2019end2}]{
    \label{fig:visualize_intra_eend}
    \begin{minipage}[c]{0.49\linewidth}
        \includegraphics[width=\linewidth]{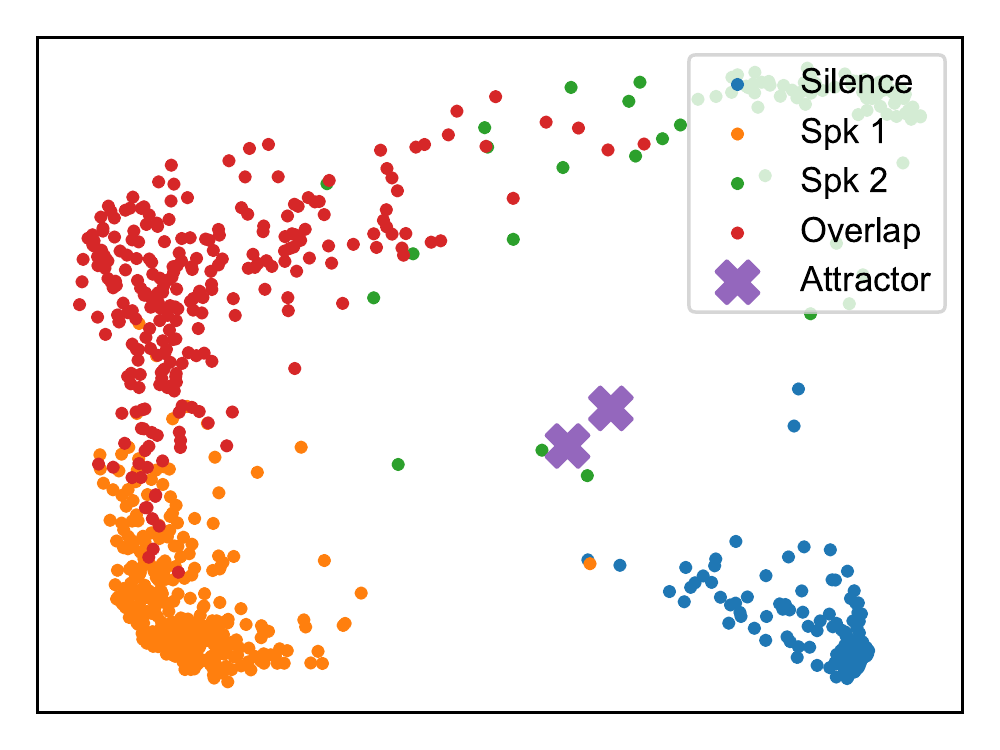}
    \end{minipage}
    \hfill
    \begin{minipage}[c]{0.49\linewidth}
        \includegraphics[width=\linewidth]{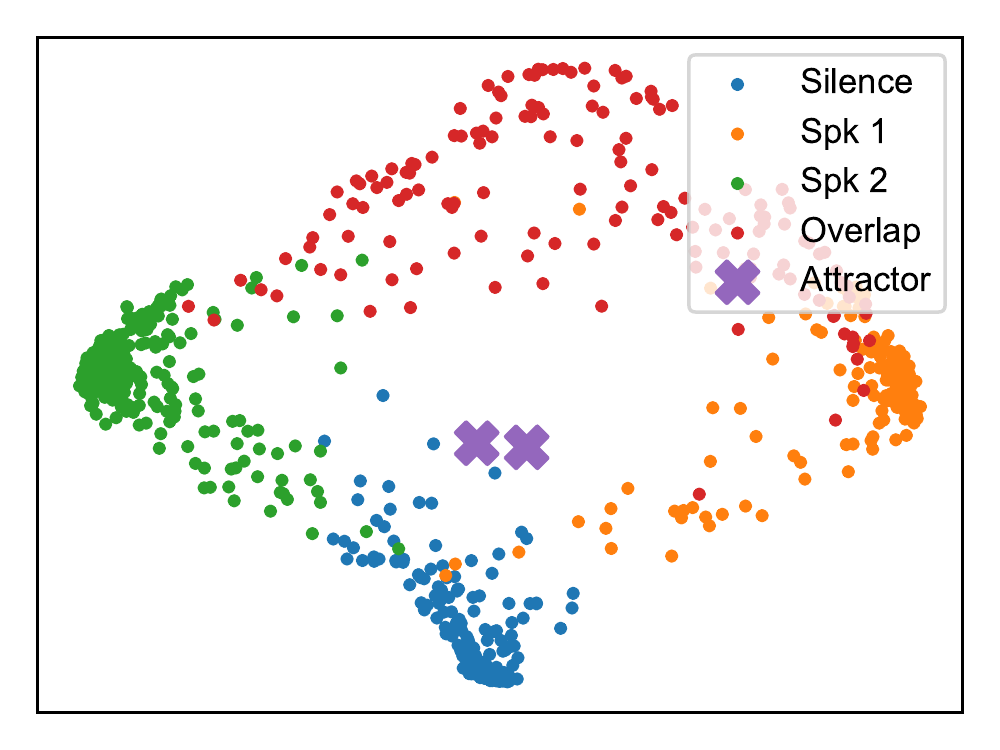}
    \end{minipage}
    }\\
    \subfloat[][EEND-EDA]{
    \label{fig:visualize_intra_eda}
    \begin{minipage}[c]{0.49\linewidth}
        \includegraphics[width=\linewidth]{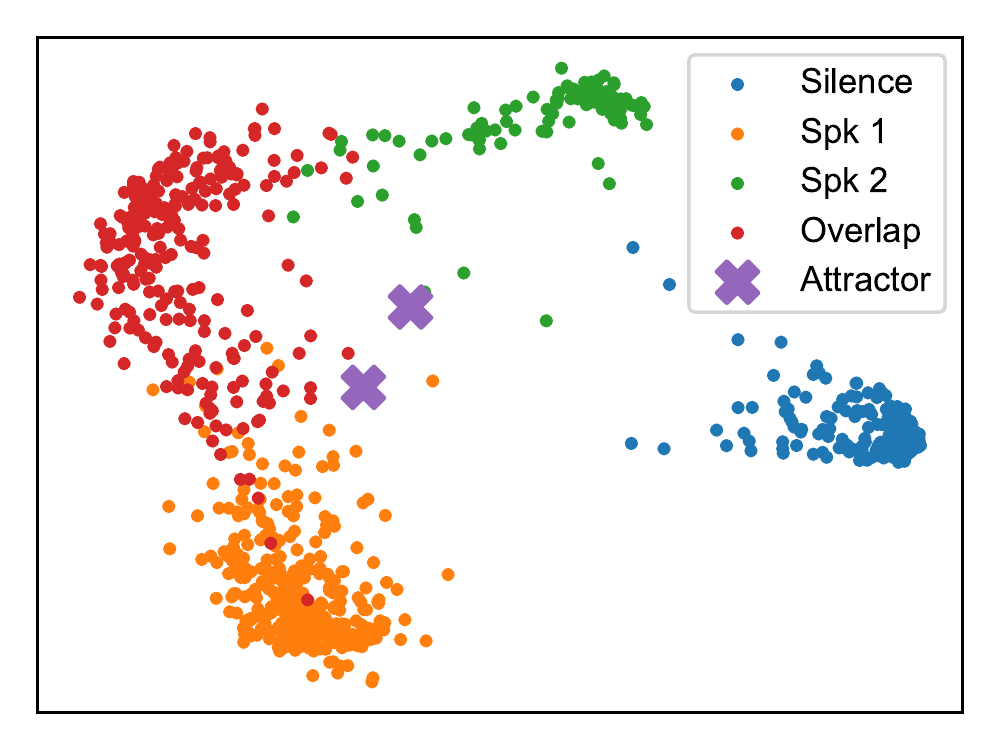}
    \end{minipage}
    \hfill
    \begin{minipage}[c]{0.49\linewidth}
        \includegraphics[width=\linewidth]{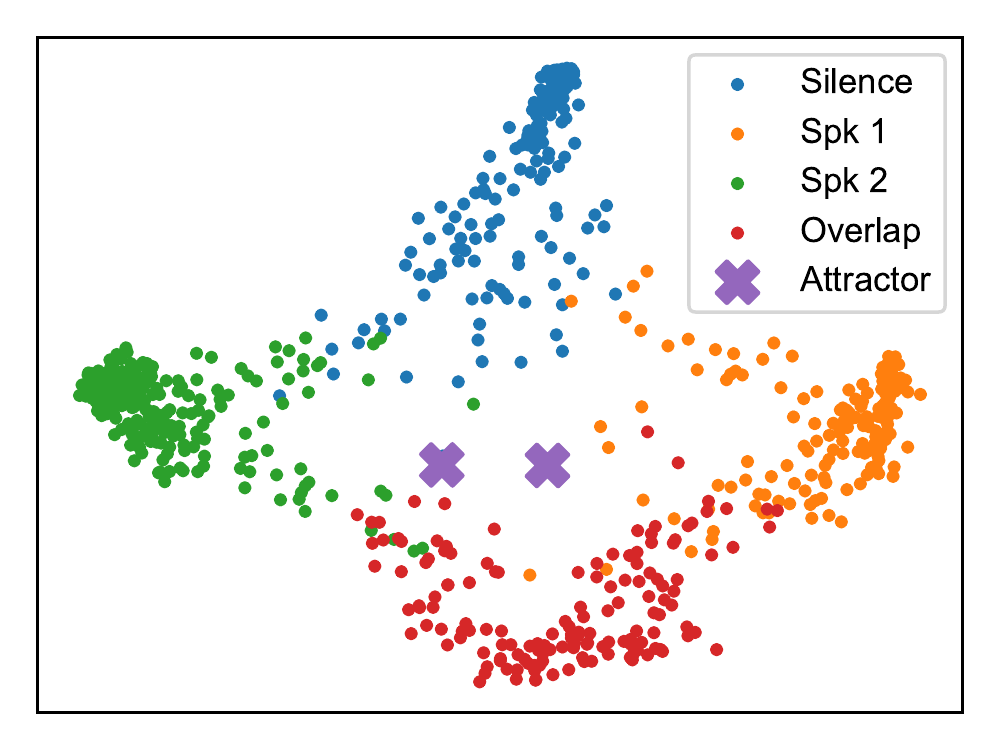}
    \end{minipage}
    }
    \caption{Visualization of embedding and attractors within each recording. For conventional EEND, weights of last fully connected layer $W_\text{cls}$ were visualized instead of attractors.}
    \label{fig:visualize_intra}
\end{figure}

For intuitive understanding of the behavior of EDA, we visualized the embeddings $\vect{e}_t$ and attractors $\vect{a}_s$ within a two-speaker mixture from Sim2spk ($\beta=2$) in \autoref{fig:visualize_intra_eda}.
They were projected to two-dimensional space by using principal component analysis (PCA).
We observed that the embeddings of two speakers were well distinguished from those of silence frames, and those of overlapped frames were distributed between the areas of the two speakers.
For EEND-EDA, two attractors were calculated for each of the two speakers successfully as in \autoref{fig:visualize_intra_eda}.
In \autoref{fig:visualize_intra_eend}, in comparison, the fixed attractors $W_\text{cls}$ of the conventional EEND were not well separated compared with the attractors calculated using EDA.

\begin{figure}[t]
    \centering
    \begin{minipage}[c]{0.49\linewidth}
        \includegraphics[width=\linewidth]{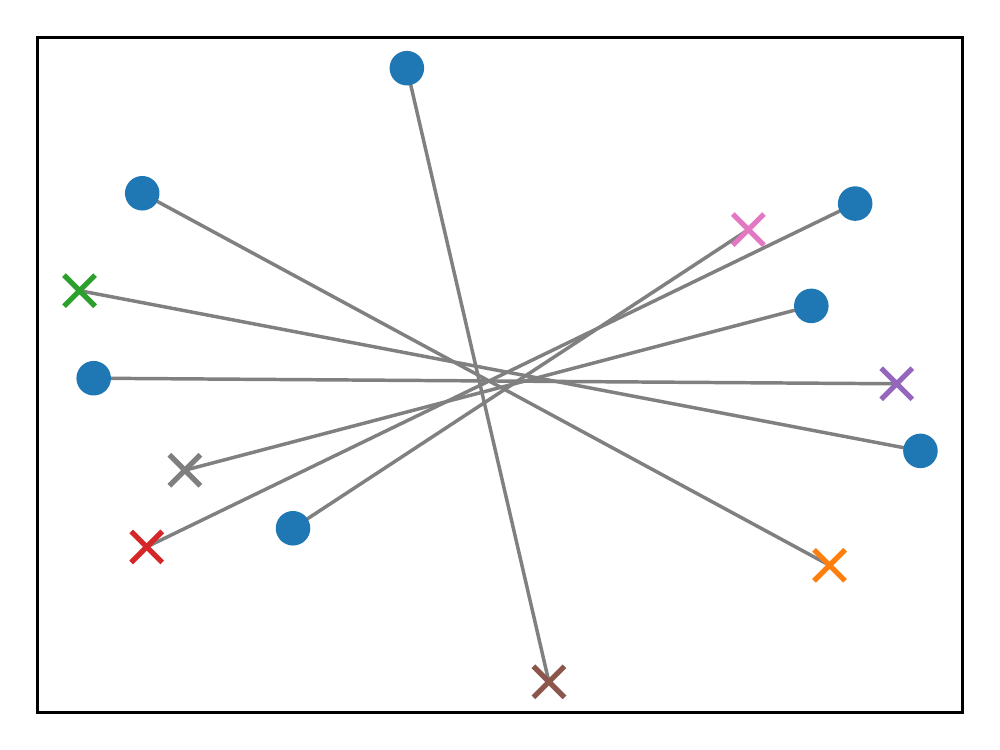}
    \end{minipage}
    \hfill
    \begin{minipage}[c]{0.49\linewidth}
        \includegraphics[width=\linewidth]{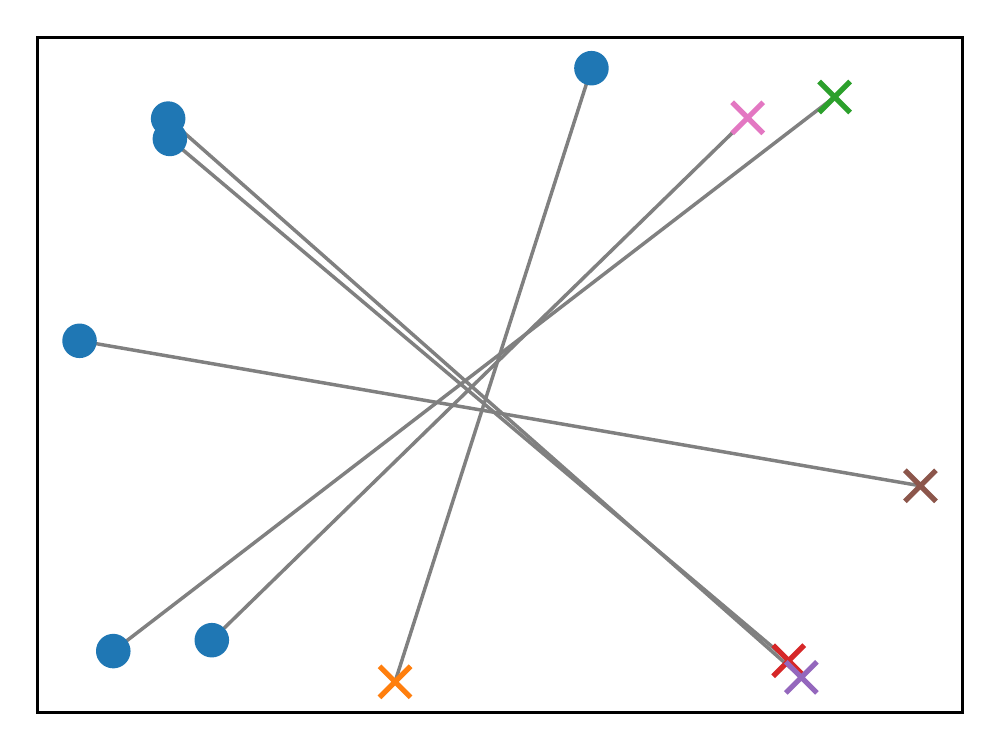}
    \end{minipage}
    \caption{Visualization of attractors across recordings. Selected speakers' attractors are marked by dots, and their interference speakers' attractors are marked by crosses. Colors of crosses correspond to speaker identities within each figure. Each pair of attractors from same mixture are connected with gray line.}
    \label{fig:visualize_inter}
\end{figure}

To understand the characteristics of attractors from EDA, we also visualized the inter-mixture relationship of attractors.
For visualization, we first chose an anchor speaker and then selected mixtures that contained the anchor speaker.
We calculated two attractors from each mixture by using EEND-EDA and mapped them onto a two-dimensional space using PCA.
The speaker assignment from the calculated attractors to speaker identifiers was based on the groundtruth labels.
\autoref{fig:visualize_inter} shows the attractors of two-speaker mixtures that contain the same anchor speaker.
It clearly shows that the each anchor speaker's attractors were not distributed near each other.

From these results, the embeddings and attractors were calculated only to separate speakers in each mixture.
We can also say that the attractors were not suited for speaker identification.
This also supports the idea that attractors are adaptively calculated from input embeddings.
A similar observation on attractors from DANet \cite{chen2017deep} in speech separation was provided in Section 5 of \cite{drude2018deep} that attractors cannot be used for speaker identification or tracing.

\subsubsection{Evaluation on the mismatched number of speakers}
We also evaluated two-speaker EEND-EDA on three-speaker datasets, and three-speaker EEND-EDA on two-speaker datasets.
We used the model trained on Sim2spk or Sim3spk for the evaluation on the simulated datasets, and used the model adapted to CALLHOME-2spk or CALLHOME-3spk for the evaluation on the real datasets.
The order of the embeddings is shuffled before being fed into EDA.
The results are shown in \autoref{tbl:results_cross_spk}.
It is clearly observed that the DERs degraded when the number of speakers during training and inference was different.
It is worth mentioning that three-speaker EEND-EDA did not work well on the two-speaker datasets; this indicates that the larger number of speakers during training does not serve the smaller number of speakers during inference.

\begin{table}[t]
    \centering
    \caption{DERs (\%) of cross evaluations of two- and three-speaker EEND-EDA. \SI{0.25}{\second} of collar tolerance was allowed.}
    \label{tbl:results_cross_spk}
    \setlength{\tabcolsep}{4.5pt}
    \resizebox{\linewidth}{!}{%
    \begin{tabular}{@{}lcccc@{}}
        \toprule
        &\multicolumn{2}{c}{Two-speaker datasets}&\multicolumn{2}{c}{Three-speaker datasets}\\\cmidrule(l{4.5pt}r{4.5pt}){2-3}\cmidrule(l{4.5pt}){4-5}
        &Sim2spk&CALLHOME&Sim3spk&CALLHOME\\
        Model&$(\beta=2)$&-2spk&$(\beta=5)$&-3spk\\\midrule
        Two-speaker EEND-EDA&2.69&8.07&28.79&20.80\\
        Three-speaker EEND-EDA&15.12&9.95&8.38&13.92\\
        \bottomrule
    \end{tabular}%
    }
\end{table}

\begin{table*}[t]
    \centering
    \caption{Step-by-step improvement on simulated datasets. For Sim2spk and Sim3spk, we used $\beta=2$ and $\beta=5$, respectively. In $\mathcal{L}_\text{exist}$ column, we show which parameters were updated using $\mathcal{L}_\text{exist}$ during training. Results on top row correspond to original setting \cite{horiguchi2020endtoend}.}
    \label{tbl:ablation_study}
    %\resizebox{\linewidth}{!}{%
    \begin{tabular}{@{}lcccccccc@{}}
        \toprule
        &&&&\multicolumn{5}{c}{Sim$k$spk}\\\cmidrule(l){5-9}
        Model&Training data&\#Epochs&$\mathcal{L}_\text{exist}$&$k=1$&$k=2$&$k=3$&$k=4$&$k=5$\\\midrule
        \multirow{4}{*}{EEND-EDA}&$k\in\{1,\dots,4\}$ &25& Update all the parameters in $f_\mathsf{EEND}$&0.39&4.33&8.94&13.76&N/A\\
        &$k\in\{1,\dots,4\}$&25& Update only $\vect{w}_\text{exist}$ and $b_\text{exist}$ &0.25&4.06&7.68&10.12&23.08\\
        &$k\in\{1,\dots,5\}$&25& Update only $\vect{w}_\text{exist}$ and $b_\text{exist}$ &0.21&4.22&8.25&10.75&13.70\\
        &$k\in\{1,\dots,5\}$&50& Update only $\vect{w}_\text{exist}$ and $b_\text{exist}$ &0.36&3.65&7.70&9.97&11.95\\
        \midrule
        \multirow{2}{*}{SA-EEND}&$k\in\{1,\dots,4\}$&50&N/A&0.60&4.39&9.40&13.56&25.22\\
        &$k\in\{1,\dots,5\}$&50&N/A&0.50&3.95&9.18&12.24&17.42\\
        \bottomrule
    \end{tabular}%
    %}
\end{table*}

\subsection{Unknown numbers of speakers}
\subsubsection{Simulated mixtures}
To train EEND-EDA to output flexible numbers of speakers' results, we finetuned the model from the two-speaker model for at most 50 epochs using Sim1spk to Sim4spk or Sim1spk to Sim5spk.
\autoref{tbl:ablation_study} shows the step-by-step improvement of the model.
Note that the results on the top row correspond to our previous paper \cite{horiguchi2020endtoend}.
First, disabling backpropagation from the attractor existence loss $\mathcal{L}_\text{exist}$ to update only $\vect{w}_\text{exist}$ and $b_\text{exist}$ improved the DERs for Sim1spk to Sim4spk.
However, we observed that the model still did not perform well on Sim5spk, which was not included in the training set.
Adding Sim5spk to the training set solved the problem as shown in the third row, which shows DERs that improved for Sim5spk from \SI{23.08}{\percent} to \SI{13.70}{\percent}.
This indicates that EEND-EDA's number of output speakers was empirically limited by its training datasets, even though it does not limit the number of output speakers with its network architecture. 
Increasing the number of training epochs further improved the DERs as shown in the last row.
We also showed the DERs computed by SA-EEND \cite{fujita2019end2} trained on a flexible number of speakers' dataset in the last two rows.
In each case, the model's output number of speakers was set to the maximum number of speakers in the dataset, \ie, four or five, and the model was trained to output null speech activities if a recording of a fewer number of speakers was input.
EEND-EDA outperformed SA-EEND in all datasets.
Hereafter, we use the EEND-EDA model of the fourth row ($k\in\{1,\dots,5\}$, 50 epochs, using $\mathcal{L}_\text{exist}$ to update only $\vect{w}_\text{exist}$ and $b_\text{exist}$ during training) and the SA-EEND model of the sixth row ($k\in\{1,\dots,5\}$, 50 epochs).

\subsubsection{CALLHOME}
\begin{table*}[t]
    \caption{DERs (\%) of CALLHOME. \SI{0.25}{\second} of collar tolerance was allowed. TDNN-based x-vector results were obtained with Kaldi recipe. DERs of single-speaker regions are reported in brackets. AHC: agglomerative hierarchical clustering, VB: Variational Bayes resegmentation \cite{diez2020analysis}, VBx: Variational Bayes HMM clustering \cite{landini2022bayesian}.}
    \captionsetup{position=top}
    \subfloat[][Results of cross-validation.]{
    \label{tbl:results_callhome_cv}
    \scalebox{0.8}{%
    \begin{tabular}{@{}lcccccccc@{}}
        \toprule
        &&\multicolumn{6}{c}{\#Speakers}\\\cmidrule(lr){3-8}
        Method&SAD&2&3&4&5&6&7&Total\\\midrule
        SA-EEND&-&8.51&19.84&26.16&36.82&48.52&38.24&19.82 (13.38)\\
        EEND-EDA &-&8.18&15.05&16.54&27.29&31.40&37.23&14.81 (8.68)\\
        \midrule
        X-vector (TDNN) + AHC&TDNN &14.66&18.42&20.46&31.40&32.62&46.43&19.48 (10.25)\\
        X-vector (TDNN) + AHC + VB&TDNN &11.68&17.22&19.71&30.24&32.07&46.49&17.80 (8.29)\\
        SA-EEND&TDNN&7.42&18.10&21.80&31.69&44.61&35.09&17.41 (10.66)\\
        EEND-EDA &TDNN &\textbf{6.79}&\textbf{13.74}&\textbf{15.53}&\textbf{25.25}&\textbf{27.65}&\textbf{34.49}&\textbf{13.36} (\textbf{7.12})\\
        \midrule
        X-vector (TDNN) + AHC&Oracle&13.68&17.04&17.89&29.96&32.55&45.20&18.04 (8.54)\\
        X-vector (TDNN) + AHC + VB&Oracle&10.94&15.85&17.40&29.23&33.97&42.69&16.57 (6.63)\\
        X-vector (ResNet101) + AHC + VBx \cite{landini2022bayesian} & Oracle  &9.83&15.23&14.29&\textbf{19.24}&\textbf{25.76}&36.25&14.21 (\textbf{4.42})\\
        SA-EEND&Oracle&6.02&16.28&20.26&30.42&43.51&35.09&15.90 (8.99)\\
        EEND-EDA & Oracle & \textbf{5.50}&\textbf{12.17}&\textbf{12.86}&23.17&27.96&\textbf{34.08}&\textbf{11.72} (5.29)\\
        \bottomrule
    \end{tabular}%
    }
    }
    \hfill
    \subfloat[][Results on CALLHOME Part 2.]{
    \centering
    \label{tbl:results_callhome2}
    \scalebox{0.8}{%
    \begin{tabular}{@{}lcc@{}}
        \toprule
        Method&SAD&DER\\\midrule
        SA-EEND &-&21.19\\
        SC-EEND \cite{fujita2020neural}&-&15.75\\
        SAD-OD-fiert SC-EEND \cite{takashima2021endtoend}&-&15.32\\
        EEND-EDA (From \cite{horiguchi2020endtoend})&-&15.29\\
        EEND-EDA &-&\textbf{12.88}\\
        \midrule
        X-vector (TDNN) + AHC&TDNN&19.43\\
        X-vector (TDNN) + AHC + VB&TDNN&17.61\\
        SA-EEND&TDNN&19.85\\
        EEND-EDA &TDNN&\textbf{13.84}\\
        \midrule
        X-vector (TDNN) + AHC&Oracle&17.02\\
        X-vector (TDNN) + AHC + VB&Oracle&15.57\\
        X-vector (ResNet101) + AHC + VBx \cite{landini2022bayesian} & Oracle &13.33\\
        SA-EEND&Oracle&16.79\\
        EEND-EDA & Oracle &\textbf{10.46}\\
        \bottomrule
    \end{tabular}%
    }
    }
\end{table*}

\begin{table}[t]
    \centering
    \caption{Confusion matrices for speaker counting on CALLHOME Part 2. X-vector-based results were obtained with oracle SAD, while EEND-based results were obtained without external SAD.}
    \label{tbl:speaker_counting}
    \captionsetup{position=top}
    \setlength{\tabcolsep}{4.5pt}
    \subfloat[][X-vector (TDNN) + AHC (Accuracy=\SI{56.4}{\percent})]{
        \begin{tabular}{@{}cc|cccccc@{}}
            \toprule
            &&\multicolumn{6}{c}{Ref. \#Speakers}\\
            &&1&2&3&4&5&6\\\midrule
            \multirow{6}{*}{\rotatebox{90}{Pred. \#Speakers}}&1 & \textbf{0}&2&1&0&0&0\\
            &2 & 0&\textbf{87}&19&3&0&0\\
            &3 & 0&59&\textbf{51}&14&3&2\\
            &4 & 0&2&4&\textbf{3}&2&1\\
            &5 & 0&0&0&0&\textbf{0}&0\\
            &6 & 0&0&0&0&0&\textbf{0}\\
            \bottomrule
        \end{tabular}
    }
    \hfill
    \subfloat[][X-vector (ResNet101) + AHC + VBx \cite{landini2022bayesian} (Accuracy=\SI{72.0}{\percent})]{
        \begin{tabular}{@{}cc|ccccccc@{}}
            \toprule
            &&\multicolumn{6}{c}{Ref. \#Speakers}\\
            &&1&2&3&4&5&6\\\midrule
            \multirow{7}{*}{\rotatebox{90}{Pred. \#Speakers}}&1 & \textbf{0}&21&3&0&0&0\\
            &2 & 0&\textbf{122}&22&2&0&0\\
            &3 & 0&3&\textbf{44}&7&0&0\\
            &4 & 0&2&5&\textbf{10}&2&1\\
            &5 & 0&0&0&1&\textbf{3}&0\\
            &6 & 0&0&0&0&0&\textbf{1}\\
            &7 & 0&0&0&0&0&1\\
            \bottomrule
        \end{tabular}
    }\\
    \subfloat[][SC-EEND \cite{fujita2020neural} (Accuracy=\SI{76.4}{\percent})]{
        \begin{tabular}{@{}cc|ccccccc@{}}
            \toprule
            &&\multicolumn{6}{c}{Ref. \#Speakers}\\
            &&1&2&3&4&5&6\\\midrule
            \multirow{6}{*}{\rotatebox{90}{Pred. \#Speakers}}&1 & \textbf{0}&1&0&0&0&0\\
            &2 & 0&\textbf{134}&20&4&0&0\\
            &3 & 0&13&\textbf{51}&10&4&2\\
            &4 & 0&0&3&\textbf{6}&1&1\\
            &5 & 0&0&0&0&\textbf{0}&0\\
            &6 & 0&0&0&0&0&\textbf{0}\\
            \bottomrule
        \end{tabular}
    }
    \hfill
    \subfloat[][EEND-EDA (Accuracy=\SI{84.4}{\percent})]{
        \begin{tabular}{@{}cc|cccccc@{}}
            \toprule
            &&\multicolumn{6}{c}{Ref. \#Speakers}\\
            &&1&2&3&4&5&6\\\midrule
            \multirow{6}{*}{\rotatebox{90}{Pred. \#Speakers}}&1 & \textbf{0}&1&0&0&0&0\\
            &2 & 0&\textbf{142}&7&1&0&0\\
            &3 & 0&5&\textbf{54}&4&0&0\\
            &4 & 0&0&13&\textbf{14}&4&1\\
            &5 & 0&0&0&1&\textbf{1}&2\\
            &6 & 0&0&0&0&0&\textbf{0}\\
            \bottomrule
        \end{tabular}
    }
\end{table}

Since the CALLHOME dataset does not include an official dev/eval split, we used the split provided in the Kaldi recipe and performed cross-validation.
For comparison with the prior work on EEND, we also report the results obtained for Part 2 of the dataset using the model adapted to Part 1.
For SAD post-processing described in \autoref{sec:sad_postprocessing}, we used the TDNN-based SAD provided in the Kaldi ASpIRE recipe\footnote{\url{https://github.com/kaldi-asr/kaldi/tree/master/egs/aspire/s5}} and oracle speech segments.

We show the number-of-speakers-wise results of cross-validation in \autoref{tbl:results_callhome_cv}.
We also show the results for only evaluated single speaker regions in brackets.
For this purpose, we chose up the most probable speakers from each time frame of the EEND-EDA results for fair comparison with x-vector-based methods.
EEND-EDA outperformed the state-of-the-art x-vector-based methods in total DERs.
One reason is that EEND-EDA can handle speaker overlap, but it showed a competitive DER (\SI{5.29}{\percent}) even when speaker overlaps were excluded from the evaluation.
Considering the number of speakers in a mixture, EEND-EDA did especially better than the x-vector-based methods with VBx clustering when the number of speakers was small (\#Speakers=2,3,4), while it was worse or on par when the number of speakers was large (\#Speakers=5,6,7).
One reason is that the pretraining was based on mixtures with at most five speakers, and another reason is that mixtures of a larger number of speakers are rare in the CALLHOME dataset.
Compared to SA-EEND, EEND-EDA achieved better DERs on all the cases.
\autoref{tbl:results_callhome2} shows the results on CALLHOME Part2.
It clearly shows that EEND-EDA outperformed the other EEND-based methods \cite{fujita2020neural,takashima2021endtoend} by over two percent of absolute DER.

\autoref{tbl:speaker_counting} shows confusion matrices for the speaker counting of x-vector (TDNN) + AHC, x-vector (ResNet101) + AHC + VBx \cite{landini2022bayesian}, SC-EEND \cite{fujita2020neural}, and EEND-EDA on CALLHOME Part 2.
Our method achieved a higher speaker counting accuracy than the other methods by a large margin.

\subsubsection{AMI headset mix}
We next evaluated our method on the AMI headset mix, which has a different domain from the pretraining data (telephone conversation vs. meeting).
We trained the model on the training set for 500 epochs and evaluated it on the dev and eval sets.
The oracle speech segments were also used for SAD post-processing.

The results are shown in \autoref{tbl:results_ami}.
EEND-EDA outperformed the x-vector-based methods on both the dev and eval sets with the oracle SAD.
Note that the x-vector-based methods tuned the PLDA parameters on the dev set, so the superiority of EEND-EDA was smaller on the dev set than the eval set.
EEND-EDA also outperformed SA-EEND with and without the oracle SAD.
We also note that the average duration of the recordings in the AMI headset mix test set is over \SI{30}{\minute}.
The performance of EEND-EDA showed that EEND-EDA generalized well to such long recordings while using \SI{200}{\second} segments during adaptation.

\begin{table}[t]
    \centering
    \caption{DERs and JERs (\%) for AMI headset mix. No collar tolerance was allowed.}
    \label{tbl:results_ami}
    \resizebox{\linewidth}{!}{%
    \begin{tabular}{@{}lccccc@{}}
        \toprule
        &&\multicolumn{2}{c}{Dev}&\multicolumn{2}{c}{Eval}\\\cmidrule(lr){3-4}\cmidrule(l){5-6}
        Method&SAD&DER&JER&DER&JER\\\midrule
        SA-EEND &-&31.66&39.20&27.70&37.50\\
        EEND-EDA & - &21.93&25.86&21.56&29.99\\\midrule
        X-vector (ResNet101) + AHC&Oracle&19.61&23.90&21.43&25.50\\
        X-vector (ResNet101) + AHC + VBx \cite{landini2022bayesian} & Oracle &16.33&\textbf{20.57}&18.99&\textbf{24.57}\\
        SA-EEND &Oracle&23.95&35.64&20.88&34.38\\
        EEND-EDA & Oracle &\textbf{15.69}&22.19&\textbf{15.80}&26.68\\
        \bottomrule
        \end{tabular}%
    }
\end{table}

\subsubsection{DIHARD II \& DIHARD III}
Finally, we evaluated our method on the DIHARD II and III datasets, which contain recordings from multiple domains.
In this evaluation, we used iterative inference with and without DOVER-Lap, each of which are described in \autoref{sec:iterative_inference} and \autoref{sec:iterative_inference_plus}, respectively, to deal with large numbers of speakers.
For SAD post-processing, we used oracle segments and the system used in the Hitachi-JHU submission to the DIHARD III challenge \cite{horiguchi2021hitachi}. 

The results are shown in Tables \ref{tbl:results_dihard2} and \ref{tbl:results_dihard3}.
We can see that iterative inference with DOVER-Lap (iterative inference+) consistently improved DERs.
Compared with the x-vector-based methods, EEND-EDA performed best on DIHARD III full, while the x-vector-based methods were better on DIHARD II and DIHARD III core.

We show the number-of-speakers-wise DERs and JERs on DIHARD III in \autoref{tbl:results_dihard3_breakdown}.
Our method performed better when the number of speakers was small and worse when the number of speakers was large.
This is why EEND-EDA performed well on DIHARD III full and worse on DIHARD II and DIHARD III eval.
We also observed that the proposed iterative inference+ improved the performance, especially in terms of JERs on a large number of speaker cases, but it was still worse than the x-vector method.
Handling a large number of speakers with EEND is left for future work.

\begin{table*}[t]
    \begin{minipage}{0.4\linewidth}
        \centering   
        \caption{DERs and JERs for DIHARD II eval. No collar tolerance was allowed.}
        \label{tbl:results_dihard2}
        \resizebox{\linewidth}{!}{%
        \begin{tabular}{@{}lccc@{}}
            \toprule
            Method&SAD&DER&JER\\\midrule
            SA-EEND & -&32.14&54.32\\
            EEND-EDA & -&29.57&51.50\\
            EEND-EDA (Iterative inference) & -&29.41&49.61\\
            EEND-EDA (Iterative inference+)& -&\textbf{28.52}&\textbf{49.77}\\
            \midrule
            X-vector (TDNN) + AHC + VBx  \cite{landini2020but}&BUT \cite{landini2020but}&\textbf{27.11}&\textbf{49.07}\\
            SA-EEND & BUT \cite{landini2020but}&32.01&54.66\\
            EEND-EDA & BUT \cite{landini2020but}&30.48&51.78\\
            EEND-EDA (Iterative inference)& BUT \cite{landini2020but}&29.80&49.99\\
            EEND-EDA (Iterative inference+)&BUT \cite{landini2020but}&29.09&50.45\\
            \midrule
            DIHARD II baseline \cite{ryant2019second}&Oracle&28.81&50.12\\
            X-vector (TDNN) + AHC + VBx \cite{landini2020but}&Oracle&\textbf{18.21}&N/A\\
            X-vector (ResNet101) + AHC  \cite{landini2022bayesian} & Oracle &23.59&43.93\\
            X-vector (ResNet101) + AHC + VBx \cite{landini2022bayesian} & Oracle &18.55&\textbf{43.91}\\
            SA-EEND & Oracle&23.25&50.30\\
            EEND-EDA & Oracle & 20.54&46.92\\
            EEND-EDA (Iterative inference)& Oracle & 21.00&45.30\\
            EEND-EDA (Iterative inference+)& Oracle & 20.24&45.62\\
            \bottomrule
        \end{tabular}%
        }
    \end{minipage}
    \hfill
    \begin{minipage}{0.58\linewidth}
        \centering
        \caption{DERs and JERs for DIHARD III eval. No collar tolerance was allowed.}
        \label{tbl:results_dihard3}
        \resizebox{\linewidth}{!}{%
        \begin{tabular}{@{}lccccc@{}}
            \toprule
            &&\multicolumn{2}{c}{Core}&\multicolumn{2}{c}{Full}\\\cmidrule(lr){3-4}\cmidrule(l){5-6}
            Method&SAD&DER&JER&DER&JER\\\midrule
            SA-EEND&-&27.49&49.64&22.64&43.14\\
            EEND-EDA & -&25.94&47.76&21.55&41.15\\
            EEND-EDA (Iterative inference) & -&25.76&45.35&21.40&39.09\\
            EEND-EDA (Iterative inference+)& -&\textbf{24.77}&\textbf{45.18}&\textbf{20.69}&\textbf{39.07}\\
            \midrule
            X-vector (TDNN) + AHC + VBx \cite{horiguchi2021hitachi} &Hitachi-JHU \cite{horiguchi2021hitachi} & \textbf{22.99}&42.44&21.48&38.73\\
            X-vector (TDNN) + AHC + VBx + OVL \cite{horiguchi2021hitachi}&Hitachi-JHU \cite{horiguchi2021hitachi}&24.58&\textbf{42.02}&21.47&\textbf{37.83}\\
            SA-EEND&Hitachi-JHU \cite{horiguchi2021hitachi}&25.79&49.20&21.29&42.68\\
            EEND-EDA & Hitachi-JHU \cite{horiguchi2021hitachi}&23.96&46.82&20.03&40.31\\
            EEND-EDA (Iterative inference)& Hitachi-JHU \cite{horiguchi2021hitachi}&24.41&44.70&20.30&38.47\\
            EEND-EDA (Iterative inference+)&Hitachi-JHU \cite{horiguchi2021hitachi}&23.43&44.93&\textbf{19.53}&38.78\\
            \midrule
            DIHARD III baseline \cite{ryant2021third}&Oracle&20.65&47.74&19.25&42.45\\
            X-vector (TDNN) + AHC + VBx \cite{horiguchi2021hitachi}&Oracle&\textbf{16.89}&38.49&15.83&34.27\\
            X-vector (TDNN) + AHC + VBx + OVL \cite{horiguchi2021hitachi}&Oracle&18.20&\textbf{38.42}&15.65&\textbf{33.71}\\
            X-vector (ResNet152) + AHC + VBx \cite{landini2021but} & Oracle &16.56&38.72&15.79&34.46\\
            SA-EEND&Oracle&20.21&46.17&16.19&39.44\\
            EEND-EDA & Oracle & 18.38&43.69&14.91&36.93\\
            EEND-EDA (Iterative inference)& Oracle & 18.87&41.58&15.21&35.08\\
            EEND-EDA (Iterative inference+)& Oracle & 17.86&41.69&\textbf{14.42}&35.30\\
            \bottomrule
        \end{tabular}%
        }
    \end{minipage}
\end{table*}

\begin{table}[t]
    \centering
    \caption{Breakdown results of DIHARD III eval for each number of speakers with oracle speech segments.}
    \label{tbl:results_dihard3_breakdown}
    \captionsetup{position=top}
    \setlength{\tabcolsep}{2.5pt}
    \subfloat[][DER (\%)]{%
    \resizebox{\linewidth}{!}{%
    \begin{tabular}{@{}lccccccccc@{}}
        \toprule
         &\multicolumn{9}{c}{\#Speakers}\\\cmidrule(l){2-10}
        Method & 1&2&3&4&5&6&7&8&9\\\midrule
        X-vector (TDNN) + AHC + VBx & \textbf{1.30}&11.43&16.76&\textbf{23.09}&\textbf{44.99}&\textbf{26.43}&\textbf{25.61}&\textbf{35.57}&\textbf{2.03}\\
        EEND-EDA & 2.80&7.52&15.79&25.63&47.66&31.73&35.47&38.19&18.73\\
        EEND-EDA (Iterative inference+)& 1.47&\textbf{6.98}&\textbf{15.55}&26.32&47.48&31.44&34.79&38.26&14.99\\
         \bottomrule
    \end{tabular}%
    }%
    }\\
    \subfloat[][JER (\%)]{%
    \resizebox{\linewidth}{!}{%
    \begin{tabular}{@{}lccccccccc@{}}
        \toprule
         &\multicolumn{9}{c}{\#Speakers}\\\cmidrule(l){2-10}
        Method & 1&2&3&4&5&6&7&8&9\\\midrule
        X-vector (TDNN) + AHC + VBx & \textbf{2.40}&16.99&44.68&\textbf{44.70}&\textbf{66.17}&\textbf{53.32}&\textbf{56.05}&\textbf{56.71}&\textbf{8.01}\\
        EEND-EDA &3.37&11.77&\textbf{38.70}&48.37&67.40&64.85&67.77&69.00&57.60\\
        EEND-EDA + iterative inference+& 3.31&\textbf{11.34}&39.60&48.76&68.46&62.41&62.65&65.36&41.23\\
         \bottomrule
    \end{tabular}%
    }%
    }
\end{table}

\section{Conclusion}
In this paper, we proposed an end-to-end speaker diarization method for unknown numbers of speakers using an encoder-decoder-based attractor calculation module called EEND-EDA.
In EEND-EDA, frame-wise embeddings are firstly calculated from an input acoustic feature sequence, then speaker-wise attractors are calculated from the embeddings using EDA, and finally diarization results are obtained by the dot product of the embeddings and attractors.
We also proposed to improve the performance of the diarization by shuffling the order of the embeddings before input to EDA and limiting the scope of backpropagation of the attractor existence loss.
To conduct fair comparisons between EEND-based methods and cascaded methods under the same SAD condition, we introduced SAD post-processing for EEND-based methods.
We also proposed iterative inference to cope with the problem of EEND-EDA's number of outputs being empirically limited by its training dataset.
The evaluations on both simulated and real datasets showed that the proposed EEND-EDA performed well in both fixed-number-of-speakers and flexible-number-of-speakers evaluations.

One possible future direction of this research is to train EEND-EDA with simulated data of a larger number of speakers. Preparing a large amount of data in advance for training increments the storage usage. Therefore, we will need a method to prepare simulated mixtures on the fly during training as recently studied in \cite{maiti2021endtoend}. In addition, to create a simulated mixture, we first create $N$ recordings each of which contains one speaker, and then mix them to be an $N$-speaker mixture. To control the overlap ratio, we increased the value of $\beta$ as the number of speakers in the mixture increased, but this leads to an increase in the duration of silence in the mixture. An investigation of a better simulation protocol is also left for future work.

Even if EEND-EDA is trained with datasets of a large number of speakers, it would still limit the maximum number of speakers by the datasets as shown in \autoref{tbl:ablation_study}.
One reason is that EEND-EDA decides the number of speakers by using a neural network trained in a fully supervised manner.
One of our later works has shown that unsupervised clustering can be introduced into EEND-EDA to remove the limitation on the output number of speakers caused by the training dataset \cite{horiguchi2021towards}.

Another direction is the network architecture.
Currently, EDA employs a vanilla LSTM encoder-decoder, but an attention-based LSTM or Transformer encoder-decoder may be possible alternatives.
Transformer encoders to extract frame-wise embeddings from input features can be also replaced with other architectures such as Conformers \cite{liu2021endtoend} or time-dilated convolutional neural networks \cite{maiti2021endtoend}.

% or


\ifCLASSOPTIONcaptionsoff
  \newpage
\fi



\bibliographystyle{IEEEtran}

\bibliography{mybib}


%\newpage


%\vfill


\end{document}